\documentclass[aps,pre,twocolumn,groupedaddress,showkeys,longbibliography]{revtex4-1}

\usepackage{amsmath,amssymb,graphicx,times}

\newcommand{\tr}{\operatorname{Tr}}
\newcommand{\re}{\operatorname{Re}}
\newcommand{\im}{\operatorname{Im}}
\renewcommand{\H}{\mathcal{H}}

\begin{document}


\title{Chebyshev-polynomial expansion of the localization length of Hermitian and non-Hermitian random chains}


\author{Naomichi Hatano}
\affiliation{Institute of Industrial Science, University of Tokyo, Komaba, Meguro, Tokyo 153-8505, Japan}
\email[]{hatano@iis.u-tokyo.ac.jp}

\author{Joshua Feinberg}
\affiliation{Department of Mathematics, University of Haifa, Mt.\ Carmel, Haifa 31905, Israel}
\email[]{joshua@physics.technion.ac.il}


\date{\today}

\begin{abstract}
We carry Chebyshev-polynomial expansion of the inverse localization length of Hermitian and non-Hermitian random chains as function of energy.
For Hermitian models, the expansion produces numerically this energy-dependent function in one run of the algorithm.
This is in strong contrast to the standard transfer-matrix method, which produces the inverse localization length for a fixed energy in each run.
For non-Hermitian models,  as in the transfer-matrix method, our algorithm computes the inverse localization length for a fixed (complex) energy.
We also find a formula of the Chebyshev-polynomial expansion of the density of states of non-Hermitian models.
As explained in more detail in the Introduction, our algorithm for non-Hermitian models may be the only available efficient algorithm for finding the density of states of models with interactions.
\end{abstract}

\pacs{72.15.Rn, 73.20.Fz}
\keywords{Anderson localization, Kernel polynomial method, Chebyshev polynomial expansion}

\maketitle

\section{Introduction}
\label{sec1}

Impurities are ubiquitous in nature and play  essential role in various physical phenomena; one of the most important phenomena is Anderson localization.
Anderson~\cite{Anderson58} had originally introduced his model  to describe localization of electrons diffusing in  randomly disordered lattices, but later on it has been applied to various other systems where waves are present in a random environment; see \textit{e.g.}\ Ref.~\cite{Lagendijk09}.
Waves, quantum mechanical or classical, that are scattered by random impurities, tend to interfere destructively with each other and  consequently become localized in space under specific conditions.

These localized waves typically have an envelope with an exponential tail
\begin{align}
|\psi(x)|\sim e^{-\kappa|x-x_0|}
\end{align}
in the limit $|x-x_0|\to \infty$, where $x_0$ is the center of localization.
The parameter  $\kappa$ specifies how strongly the wave is localized, and is referred to as the {\em inverse localization length}.
A small value of $\kappa$ indicates a weakly localized (\textit{i.e.}, widely spread) wave, while a large value a strongly localized (\textit{i.e.}, narrowly spread) one.

The inverse localization length $\kappa$ of each eigenstate of the Hamiltonian depends on its eigenenergy.
For large quantum-mechanical systems, the function $\kappa(E)$ is typically larger (stronger localization) near energy-band edges, smaller (weaker localization) away from them, and may even vanish in an energy range.
It is now widely accepted that  in one dimension, in the absence of inter-particle interactions, almost all eigenstates are localized, that is, $\kappa(E)>0$ for any $E$,  while in three spatial dimensions there is an energy region where $\kappa(E)=0$, namely a phase of extended states, with a transition to a phase of localized states at the so-called {\em mobility edge}. It has been stressed~\cite{Kawarabayashi96,Slevin97,Slevin99,Slevin01} that a detailed finite-size scaling analysis of numerical data is essential to fix accurately the exponent $\nu$, which determines the energy dependence  $\kappa(E)\sim |E-E_c|^\nu$ of the inverse localization length in the localized phase, as one approaches the mobility edge $E_c$, and thereby the universality class of the Anderson localization transition.
Numerical methods for computing the inverse localization length precisely are still in need for more intricate systems, such as quantum Hall systems, systems with spin-orbit coupling, and topological insulators.

In the present paper, we introduce a novel numerical method for computing the energy dependence of the inverse localization length $\kappa(E)$;
we derive an expansion of the function $\kappa(E)$ in terms of Chebyshev polynomials $T_n(E)$. The most popular method at present is presumably to find it as the Lyapunov exponent of the random transfer matrix~\cite{Ishii73,Kawarabayashi96,Slevin97,Slevin99,Slevin01}. Our new method is completely different; it extracts the inverse localization length directly from the density of states of the Hamiltonian.
The most prominent practical difference may be the following point. Each run of the transfer matrix method finds the inverse localization length for a very large system at a fixed energy. In contrast, each run of our method finds $\kappa(E)$ \textit{as a function} for a moderately large system.

We also present an algorithm for computing the inverse localization length of non-Hermitian Hamiltonians in the complex energy plane, using again the Chebyshev-polynomial expansion.
Non-Hermitian Hamiltonians and Liouvillians can appear in quantum mechanics when the environment is traced out in open quantum systems~\cite{Livshitz56,Feshbach58,Feshbach62,Rotter91,Rotter09,Sasada11,Hatano13,Hatano14,Petrosky96,Petrosky97}.
Interest in non-Hermitian quantum mechanics was renewed in 1990's, when several important studies on non-Hermitian Hamiltonians appeared, including a non-Hermitian extension of a model of the Anderson localization~\cite{HN96,HN97,Hatano98} and the $\mathcal{PT}$-symmetric quantum mechanics~\cite{Bender98,Bender99,Bender02}.
In the former, competition between randomness and non-Hermiticity was found;
the inverse localization length vanishes at the critical value of a non-Hermitian parameter.
If we go further away from the realm of quantum mechanics, the presence of randomness in non-Hermitian matrices is quite common, such as in the Fokker-Planck dynamics~\cite{Chalker97,Chalker00}, fluid dynamics~\cite{Giona04} and in neural networks~\cite{AHN15,Ahmadian15}, and the concept of the Anderson localization flourishes.
Our method should come in handy in analyzing such non-Hermitian systems.

Chebyshev-polynomial algorithms, and more generally the kernel-polynomial methods, employ repeated multiplication of some vector by the Hamiltonian matrix.
Consequently, for sparse Hamiltonian matrices (which is the case for nearest-neighbor hopping systems), it is possible to carry out the algorithms by storing in computer memory only several vectors of the size of the Hilbert space.
Thanks to this feature, Chebyshev-polynomial algorithms as a matter of principle, should be more efficient in analyzing many-body Hamiltonians with interactions, which typically involve Hilbert spaces of very large dimensions.
The present algorithm for non-Hermitian models is, as far as we know, potentially the only one available for finding the density of states of general large non-Hermitian models, particularly large many-body ones, as a function of complex energy;
we are only aware of existing algorithms that find individual complex eigenvalues~\cite{Jang95,Fernandez99,Keller99,Moiseyev98,Aoyama06,Myo14,Landau16}  and those that find spectra of specific types of large sparse non-Hermitian matrices~\cite{Rogers09,Metz11,Neri12}.

The paper is organized as follows.
We first review in Section~\ref{sec2.1} the Chebyshev-polynomial expansion of the density of states $\rho(E)$ of Hermitian Hamiltonians.
We then move on to our new Chebyshev-polynomial expansion of the inverse localization length $\kappa(E)$ in Section~\ref{sec2.2}.
We present in Section~\ref{sec2.3} two demonstrations of the method, employing the random-potential and random-hopping tight-binding models.

We turn to our method for the inverse localization length of non-Hermitian Hamiltonians in Section~\ref{sec3}.
After derivation of the expansion formula, we demonstrate it in the case of a non-Hermitian random-sign model~\cite{FZ99}.

We finally present the method for the density of states of non-Hermitian Hamiltonians in Section~\ref{sec4}.
A demonstration with full (\textit{i.e.}, non-sparse) non-Hermitian random matrices follows.

The methods for the density of states given in Sections~~\ref{sec2.1} and~\ref{sec4} do not depend on the dimensionality of the system.
The methods for the inverse localization length given in Sections~\ref{sec2.2} and~\ref{sec3} are primarily for one-dimensional systems, because they utilize the Thouless formula. 
Some comments on the applicability of our method to higher spatial dimensional systems are deferred to Sec.~\ref{sec2.2} below.

\section{The Chebyshev-polynomial method for disordered Hermitian chains}
\label{sec2}

\subsection{Computing the density of states}
\label{sec2.1}

Let us first briefly overview the Chebyshev-polynomial expansion method for computing the density of states of a (Hermitian) Hamiltonian matrix~\cite{Silver94,Silver96,Silver97}.
The method in which errors due to termination of the expansion are taken care of is often called the {\em kernel-polynomial method}~\cite{Weisse06}.
Throughout this paper, for the sake of simplicity, we employ the simpler, straightforward Chebyshev-polynomial expansion, and avoid any issues of optimized truncation for minimizing errors pertaining to the kernel-polynomial method. We justify this simplification by demonstrating numerically the convergence of the expansion as the number of terms summed increases. The method works most efficiently for large sparse matrices, to which point we will come back below.
{\em We emphasize that the discussion in the present sub-section is not restricted to one-dimensional systems.}

Suppose that an $L\times L$ Hermitian matrix $H$ has the real (unknown) eigenvalues $\{E_\nu|\nu=1,2,\cdots,L\}$.
Its density of states is given by
\begin{align}\label{eq10}
\rho(E):=\frac{1}{L}\sum_{\nu=1}^L\delta(E-E_\nu)\,.
\end{align}
For large systems $L\gg 1$, the sum over the dense delta-function spikes in Eq.~\eqref{eq10} is typically smoothed out into a continuous function $\rho(E)$. 
The strategy is to expand the density of states $\rho(E)$ in terms of a set of orthogonal polynomials;
often used are Chebyshev polynomials. To employ Chebyshev polynomials, we have to rescale the matrix $H$ such that all its eigenvalues $E_\nu$ lie in the range $[-1,1]$, which is the standard domain of orthogonality of the Chebyshev polynomials. In order to determine the required scaling factor, the upper and lower bounds of the eigenvalue spectrum are roughly estimated \textit{e.g.}\ by the Gershgorin circle theorem. We assume that the matrix $H$ has been already normalized properly throughout this and next subsections.

Chebyshev polynomials of the first kind, which are defined on $[-1,1]$ by 
\begin{align}\label{eq20}
T_n(x):=\cos(n \arccos x),
\end{align}
constitute a set of orthogonal polynomials that satisfy the orthogonality relation 
\begin{align}\label{eq30}
\int_{-1}^1T_n(x)T_m(x)\frac{dx}{\sqrt{1-x^2}}=
\begin{cases}
0&\quad\mbox{for $n\neq m$,}\\
\pi&\quad\mbox{for $n=m=0$,}\\
\pi/2&\quad\mbox{for $n=m>0$}
\end{cases}
\end{align}
and the three-term recursion relation 
\begin{align}\label{eq40}
T_{n+1}(x)=2xT_n(x)-T_{n-1}(x)\,,
\end{align}
with $T_0(x)=1$ and $T_1(x)=x$. These polynomials have definite parity $T_n(-x) = (-1)^nT_n(x)$. 

We can use this complete set of orthogonal polynomials to expand the density of states in the form
\begin{align}\label{eq100}
\rho(E)=\frac{1}{\sqrt{1-E^2}}\sum_{n=0}^\infty \mu_n T_n(E)\,.
\end{align}
The orthogonality relation~\eqref{eq30} determines the expansion coefficients as
\begin{align}\label{eq110}
\mu_n&=\frac{2}{\pi}\int_{-1}^1T_n(E)\rho(E)dE
\end{align}
for $m>0$ along with
\begin{align}
\mu_0&=\frac{1}{\pi}\int_{-1}^1T_0(E)\rho(E)dE
=\frac{1}{\pi}\int_{-1}^1 \rho(E)dE=\frac{1}{\pi}.
\end{align}
Substituting the density of states~\eqref{eq10} into the expression~\eqref{eq110}, we have
\begin{align}\label{eq120}
\mu_n=\frac{2}{\pi}\frac{1}{L}\sum_{\nu=1}^LT_n(E_\nu)=\frac{2}{\pi}\frac{1}{L}\tr T_n(H),
\end{align}
where we used 
\begin{align}\label{eq130}
\sum_{\nu=1}^L{E_\nu}^k=\tr H^k.
\end{align}
The expansion~\eqref{eq100}  can be therefore rewritten as
\begin{align}\label{eq140}
\rho(E)=\frac{1}{\pi\sqrt{1-E^2}}\left[1+2\sum_{n=1}^\infty\left( \frac{1}{L}\tr T_n(H)\right)T_n(E)\right].
\end{align}

The key aspect of the method is the fact that we can evaluate the expansion coefficients by numerical matrix multiplication.
Using the diagonalizing basis of the matrix $H$, we can show that the matrix polynomial $T_n(H)$ should satisfy the recursion relation of the same form as Eq.~\eqref{eq40}:
\begin{align}\label{eq145}
T_{n+1}(H)=2HT_n(H)-T_{n-1}(H)
\end{align}
with $T_0(H)=I$, which is the $L\times L$ identity matrix, and $T_1(H)=H$.
We can therefore recursively generate the matrix polynomial $T_n(H)$ by matrix multiplications (once every order), and thereby find the expansion coefficient $\frac{1}{L}\tr T_n(H)$ in Eq.~\eqref{eq140}.

In practical numerical calculation, we of course have to truncate the summation over $n$ in the expansion~\eqref{eq140}.
Since all the  $N$ real roots of $T_N(E)$ lie in the domain $[-1,1]$, the Chebyshev polynomial of the $N$th order accounts for oscillations of wavelength $\sim 1/N$.
This implies that the truncation at the $N$th order can reproduce structures up to the resolution of order $1/N$. 
On the other hand, the level spacing is of order $1/L$ for a matrix with an almost uniform density of states, which means that the truncation at the order $N\gtrsim L$ is legitimate for finding the general features of the density of states. Various other methods for minimizing truncation errors have also been devised~\cite{Weisse06};
as was mentioned above, for simplicity of discussion and demonstration, we do not employ any such methods throughout the present paper.

As another comment, any kind of orthogonal polynomial should work in the expansion, but  Chebyshev polynomials usually work best in practical numerical calculations.
We use only Chebyshev polynomials in the present paper.
For problems with unbounded spectrum, \textit{e.g.}, for a random model with the Cauchy (Lorentzian) distribution, we would need orthogonal polynomials with infinite support, \textit{e.g.}\ the Hermite polynomials, although it is typically less stable numerically because the expansion coefficients are often less convergent.

When one applies the present algorithm to a many-body Hamiltonians with interactions, the Hamiltonian matrix can be sparse but very large.
For such matrices, the trace operation in Eq.~\eqref{eq140} is often replaced by Monte Carlo summation over a set of basis vectors less than $L$~\cite{Silver94,Silver96,Silver97}.
We can then carry out the algorithm by storing only a few vectors in the computer memory rather than storing the whole matrix;
furthermore, multiplying a vector by a sparse $L\times L$ matrix only takes CPU time of order $L$.
This is the advantage of the Chebyshev-polynomial method over the diagonalization of the matrix itself (which consumes memory size of order $L^2$ and CPU time of order $L^3$).
For the sake of demonstration, however, throughout this paper, we use only matrices that we can store in the computer memory.

\subsection{Chebyshev-polynomial expansion of the inverse localization length}
\label{sec2.2}

We now introduce the Chebyshev-polynomial expansion of the inverse localization length $\kappa(E)$ of the Hermitian {\em one-dimensional} random tight-binding model, whose Hamiltonian is given by
\begin{align}\label{tight}
H&=-\sum_{x=1}^{L-1} (t^*_{x,x+1} |x+1\rangle\langle x|+t_{x,x+1}|x\rangle\langle x+1|) 
\nonumber\\
&+ \sum_{x=1}^L V_x |x\rangle\langle x|\,.
\end{align}
The idea is simple;  we employ the Thouless formula~\cite{Thouless72}, which relates the inverse localization length $\kappa(E)$ to the density of states $\rho(E)$ in the form
\begin{align}\label{eq150}
\kappa(E)=\int_{-1}^1 \rho(x)\ln|E-x|dx-\ln|\tau|,
\end{align}
where $|\tau|$ is the geometric mean of the moduli of the hopping amplitudes, and then expand the density $\rho(E)$ according to Eq.~\eqref{eq100}.

We can sketch the derivation of the Thouless formula as follows. 
The end-to-end Green's function $G_{1L}(E)$ of a chain of length $L$ under open boundary conditions is given by
\begin{align}\label{eq155}
G_{1L}(E)&=\frac{\displaystyle \prod_{x=1}^{L-1}t_{x,x+1}}{\displaystyle \det(E-H)}
=\frac{\displaystyle \prod_{x=1}^{L-1}t_{x,x+1}}{\displaystyle \prod_{\nu=1}^L (E-E_\nu)},
\end{align}
while it presumably decays as $\exp(-\kappa(E)L)$.
Taking logarithm of the moduli of  both sides of Eq.~\eqref{eq155} results in the formula~\eqref{eq150}, with $\ln |\tau|$ being the average of $\ln |t_{x,x+1}|$ over the $L$ sites.

We note that attempts were made to extend the Thouless formula beyond one spatial dimension~\cite{Canisius85}.
The Thouless formula~\eqref{eq150} relates, essentially, the inverse localization length and the trace over the logarithm of the characteristic polynomial of the tight binding hopping Hamiltonian, $\tr\log(E-H)$. 
The latter trace may be represented straightforwardly as an integral over the eigenvalue density (see Eq.~\eqref{eq150}).  
The inverse localization length encodes directional information about the spatial behavior of the associated wave function. 
The purely spectral quantity $\tr\log(E-H)$, on the other hand, being independent of any particular basis in the Hilbert space, does not express any such spatial information. 
The fact that these two quantities can be related is peculiar only to one spatial dimension, since the energy eigenstate has only one direction to decay or grow along. 
Such a connection between the inverse localization length and $\tr\log(E-H)$ cannot hold, therefore, in higher spatial dimensions. 
The authors of Ref.~\cite{Canisius85} argued that $\tr\log(E-H)$ may contain approximate qualitative information about localization in higher spatial dimension, but this claim seems dubious.

After inserting the Chebyshev-polynomial expansion of the density of states Eq.~\eqref{eq100} into the Thouless formula~\eqref{eq150} we obtain
\begin{align}\label{eq160}
\kappa(E)=\sum_{n=0}^\infty \mu_n f_n(E) -\ln |\tau|,
\end{align}
where
\begin{align}\label{eq170}
f_n(E):=\int_{-1}^1 T_n(x)\ln|E-x|\frac{dx}{\sqrt{1-x^2}}
\end{align}
(with $|E|\leq 1$, by assumption).

Following the results of Appendix~\ref{appA} we obtain
\begin{align}\label{eq180}
f_n(E)&=-\frac{\pi}{n}T_n(E)
\end{align}
for $n>0$, along with
\begin{align}
f_0(E)&=-\pi\ln2.
\end{align}
We thereby arrive at the expansion
\begin{align}\label{eq210}
\kappa(E)&=-2\sum_{n=1}^\infty\frac{\frac{1}{L}\tr T_n(H)}{n}T_n(E)-\ln(2|\tau|).
\end{align}
Note the resemblance of the factor $1/n$ in Eq.~\eqref{eq210} to the one appearing in the Taylor expansion of logarithm $\ln(1-x)$.
Indeed, simple substitution of Eq.~\eqref{eq10} into Eq.~\eqref{eq150} results in
\begin{align}\label{eq211}
\kappa(E)&=
\frac{1}{L}\sum_{\nu=1}^L \ln|E-E_\nu| -\ln|\tau|
\nonumber\\
&=\frac{1}{L}\tr \ln|E-H|-\ln|\tau|,
\end{align}
from which we can derive the expansion~\eqref{eq210} directly by expanding the logarithm according to Eq.~\eqref{eqA90} in Appendix~\ref{appA}.
Thanks to the suppressing factor $1/n$ in the expansion coefficient, higher-order Chebyshev polynomials contribute less in the expansion of $\kappa(E)$ than in the one of $\rho(E)$, and hence the former is generally smoother than the latter, as we will demonstrate below.
This paper presents the Chebyshev-polynomial expansion of the inverse localization length for the first time, as far as we know.
 
\subsection{Numerical Demonstration}
\label{sec2.3}

\begin{figure}
\includegraphics[width=0.4\textwidth]{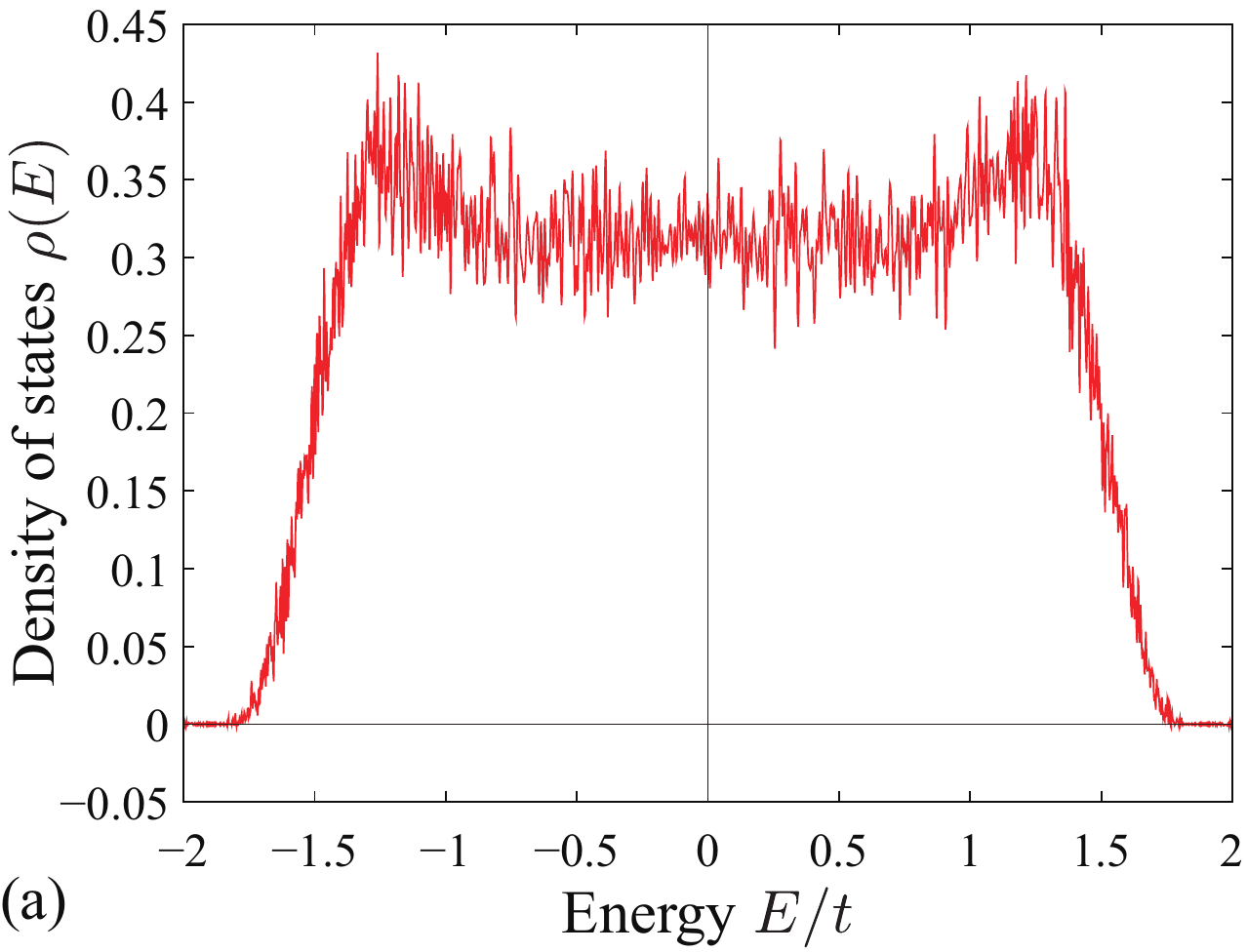}
\vspace{\baselineskip}
\\
\includegraphics[width=0.4\textwidth]{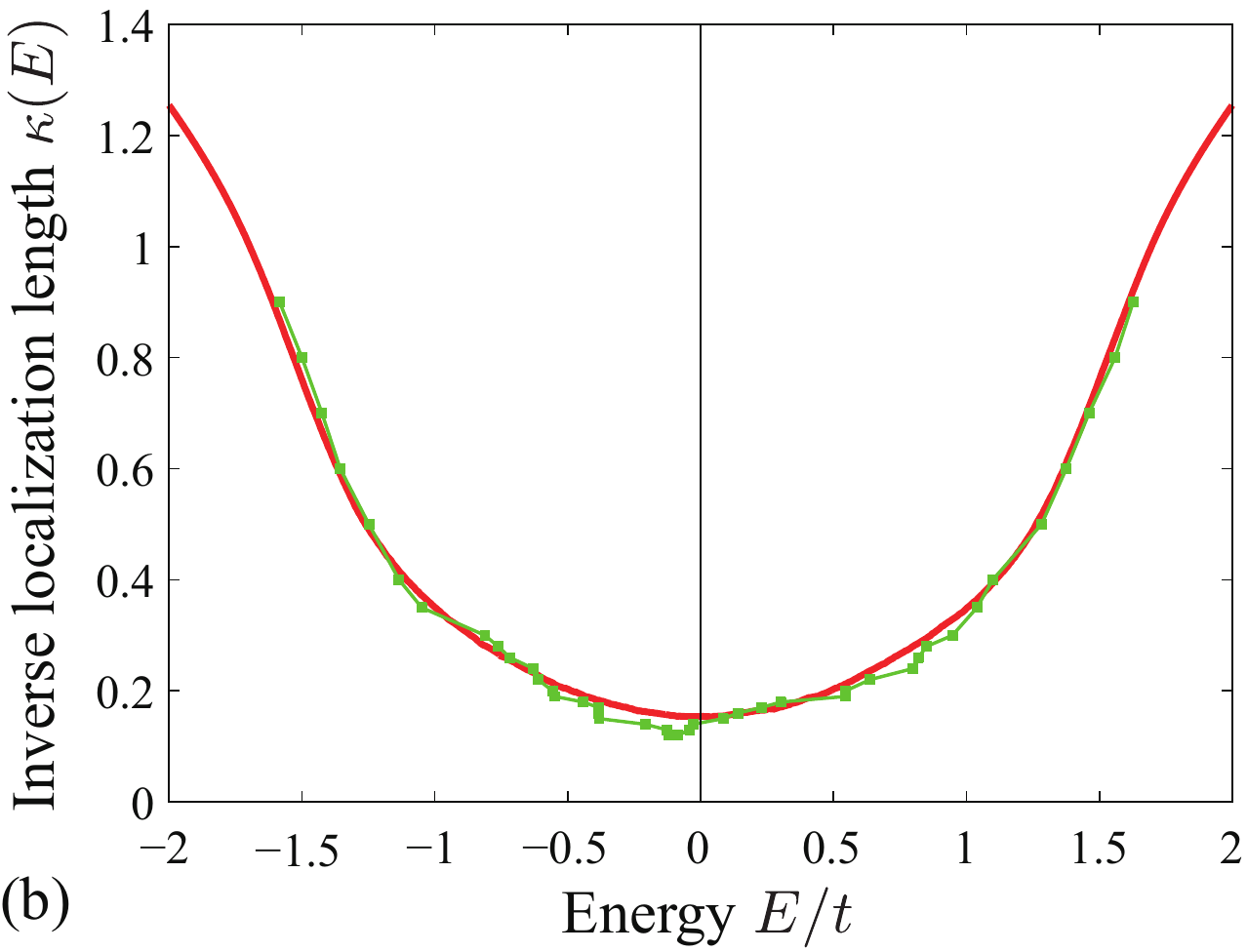}
\vspace{\baselineskip}
\\
\includegraphics[width=0.4\textwidth]{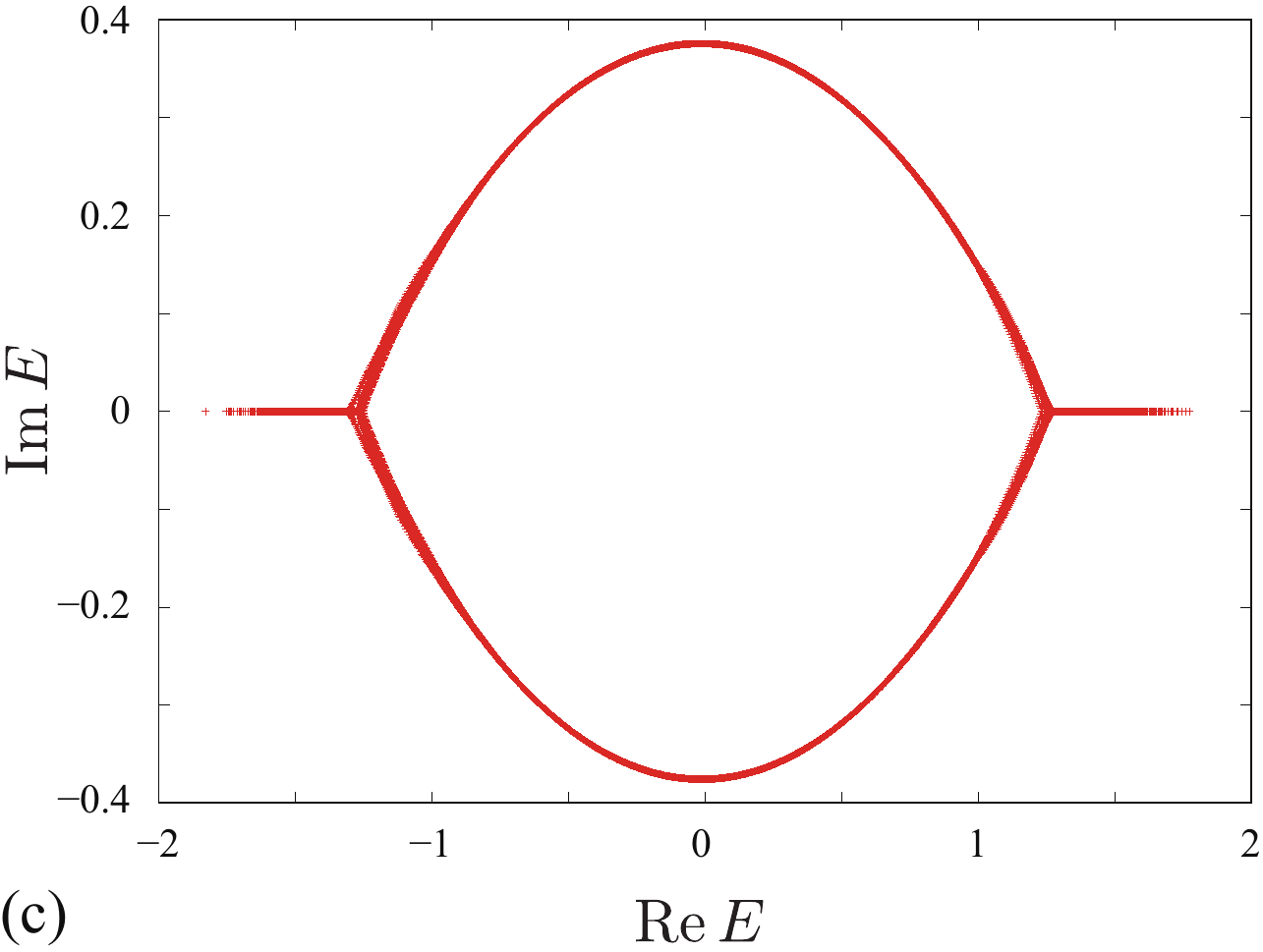}
\caption{(a) The density of states and (b) the inverse localization length computed from the Chebyshev-polynomial expansion up to the 1000th order of the random-potential chain~\eqref{eq200-1} of length $L=1001$ under open boundary conditions.
We averaged over a set of 1000 random samples. For comparison, we also plotted in (b) in green dots the inverse localization length given in Fig.~14 of Ref.~\cite{HN97}, which was deduced (see the text for more details) by monitoring the changes in the spectrum as we increase the real parameter $g$ for one sample of the HN model~\eqref{eq210-1} of length $L=1000$ under periodic boundary conditions. (c) An example of the spectrum of the model~\eqref{eq210-1} of length $L=10000$ for $g=0.5$ with the random potential in the range $[-1,1]$.
In all panels, we have removed the normalization of the spectrum into the region $[-1,1]$ and use the original energy scale with $t=1$.}
\label{fig1}
\end{figure}

Let us demonstrate the Chebyshev-polynomial expansion of the density of states, Eq.~\eqref{eq140}~\cite{Silver94,Silver96,Silver97}, as well as that of the inverse localization length, Eq.~\eqref{eq210}, for random-potential and random-hopping tight-binding models.
In Figs.~\ref{fig1} and~\ref{fig3} below in the present subsection, we remove the normalization of the spectrum into the region $[-1,1]$ and show the plots in the original energy scale.

First, Fig.~\ref{fig1} shows the results for the random-potential model
\begin{align}\label{eq200-1}
H=-\frac{t}{2}\sum_{x=1}^{L-1}\left(|x+1\rangle\langle x|+|x\rangle\langle x+1|\right)
+\sum_{x=1}^L V_x|x\rangle\langle x|\,.
\end{align}
(Here we have set $t_{x,x+1} = t/2$ in Eq.~\eqref{tight}.) We sampled the potential $V_x$ at each site randomly from the uniform distribution on $[-1,1]$, taking $t$ as the unit of energy. In both Fig.~\ref{fig1}~(a) and~(b), we computed the arithmetic average over the same set of 1000 random samples of length $L=1001$ under open boundary conditions and terminated the Chebyshev-polynomial expansion at the 1000th order.

\begin{figure}
\includegraphics[width=0.42\textwidth]{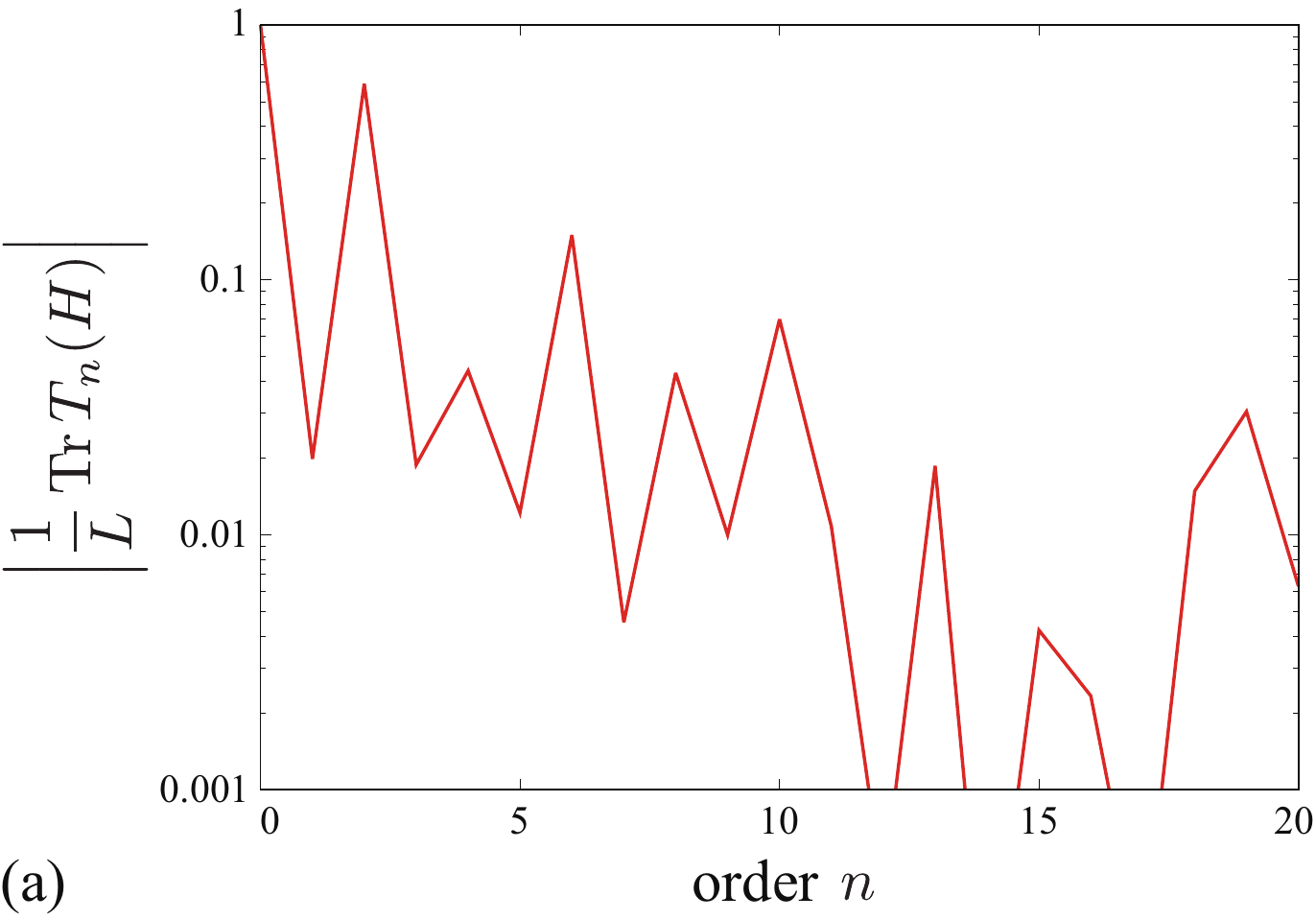}
\vspace{\baselineskip}
\\
\includegraphics[width=0.42\textwidth]{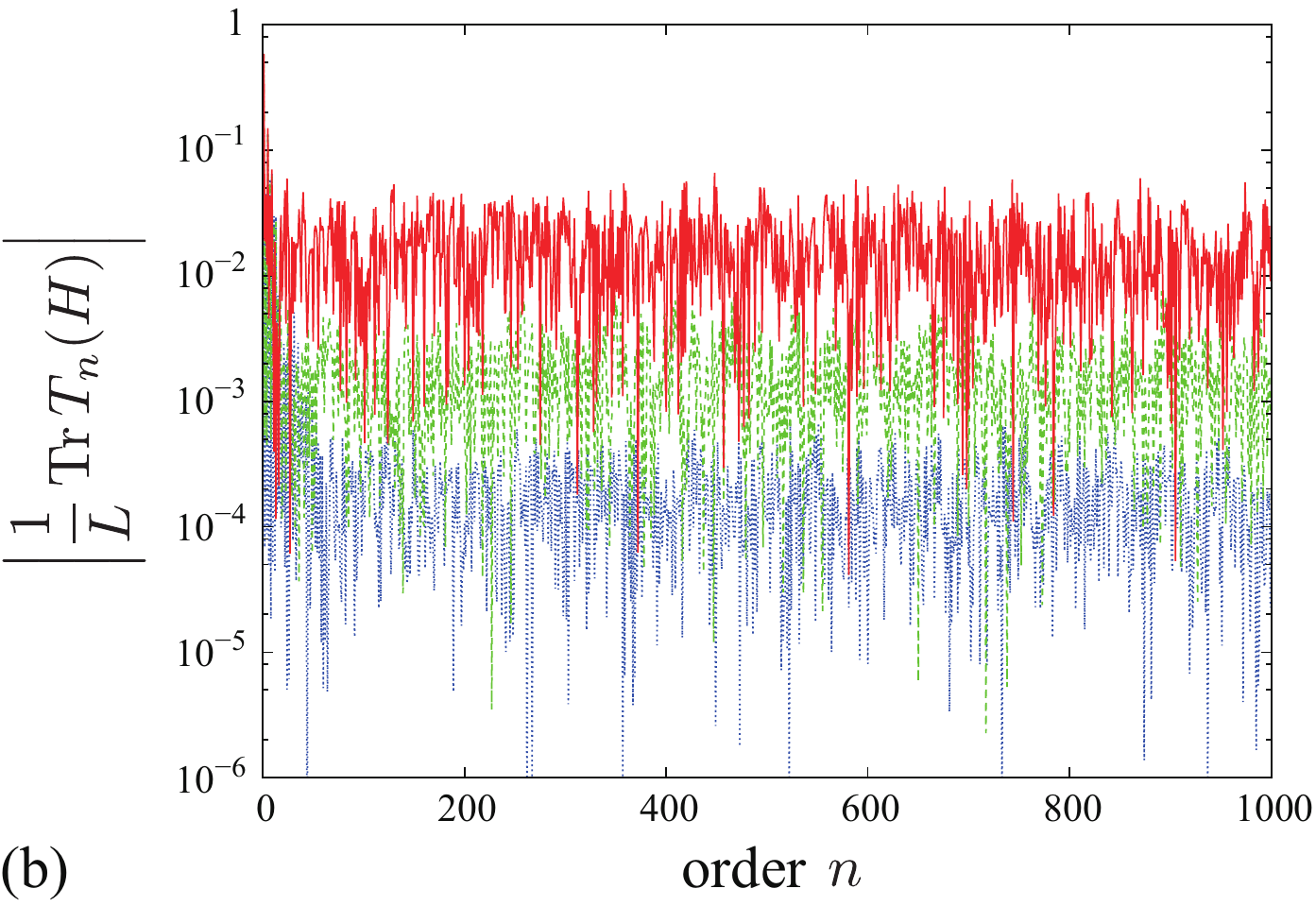}
\caption{The modulus of $\frac{1}{L}\tr T_{n}(H)$ from the zeroth order (unity) (a) to the 20th order and (b) to the 1000th order.
For (a), we used only one random sample of the random-potential chain~\eqref{eq200-1} of length $L=1001$, whereas for (b), we plotted the data for one sample (solid red line), the average over 100 samples (broken green line) and for 10 008 samples (dotted blue line).}
\label{fig2}
\end{figure}

\begin{figure}
\includegraphics[width=0.4\textwidth]{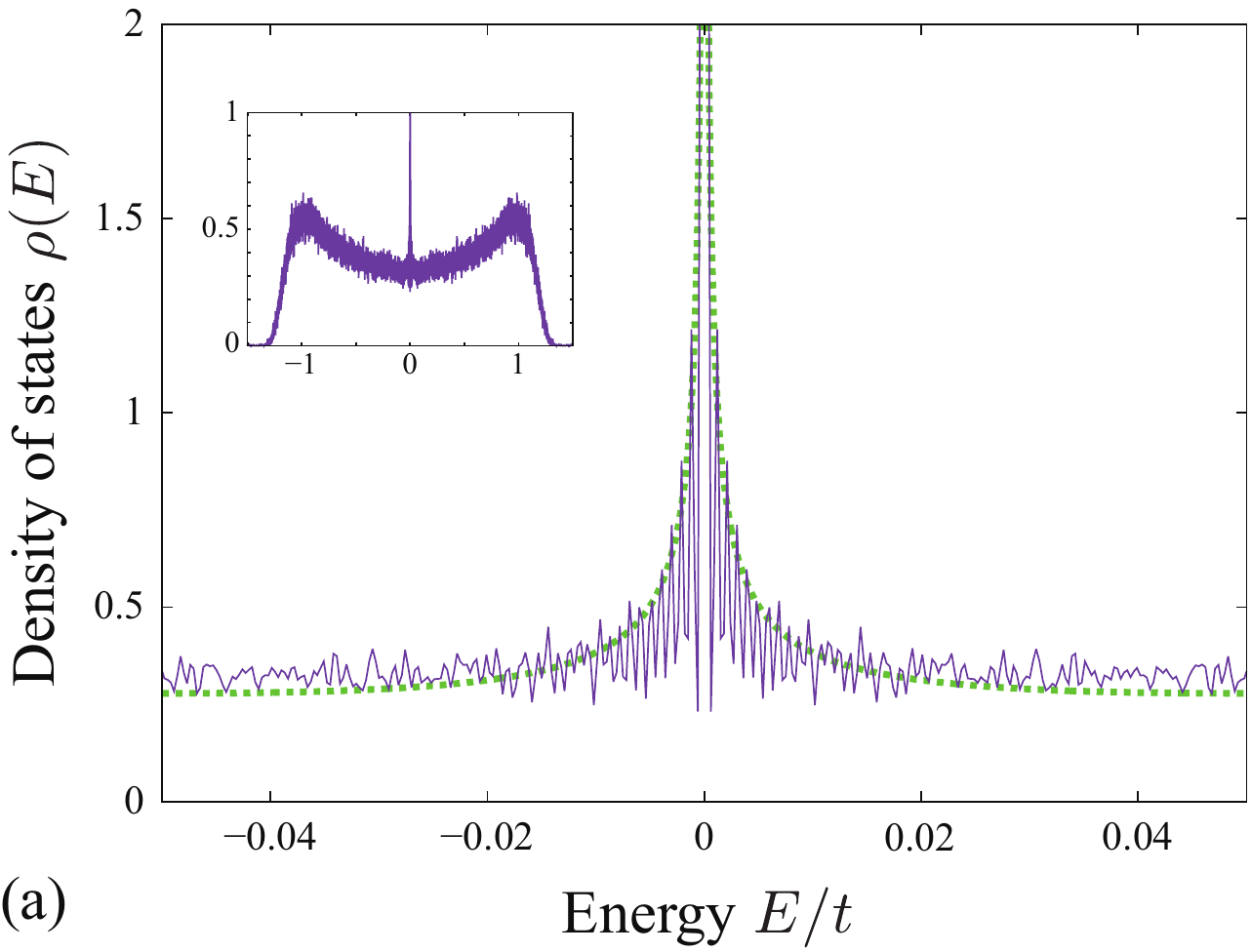}
\\
\vspace{\baselineskip}
\includegraphics[width=0.4\textwidth]{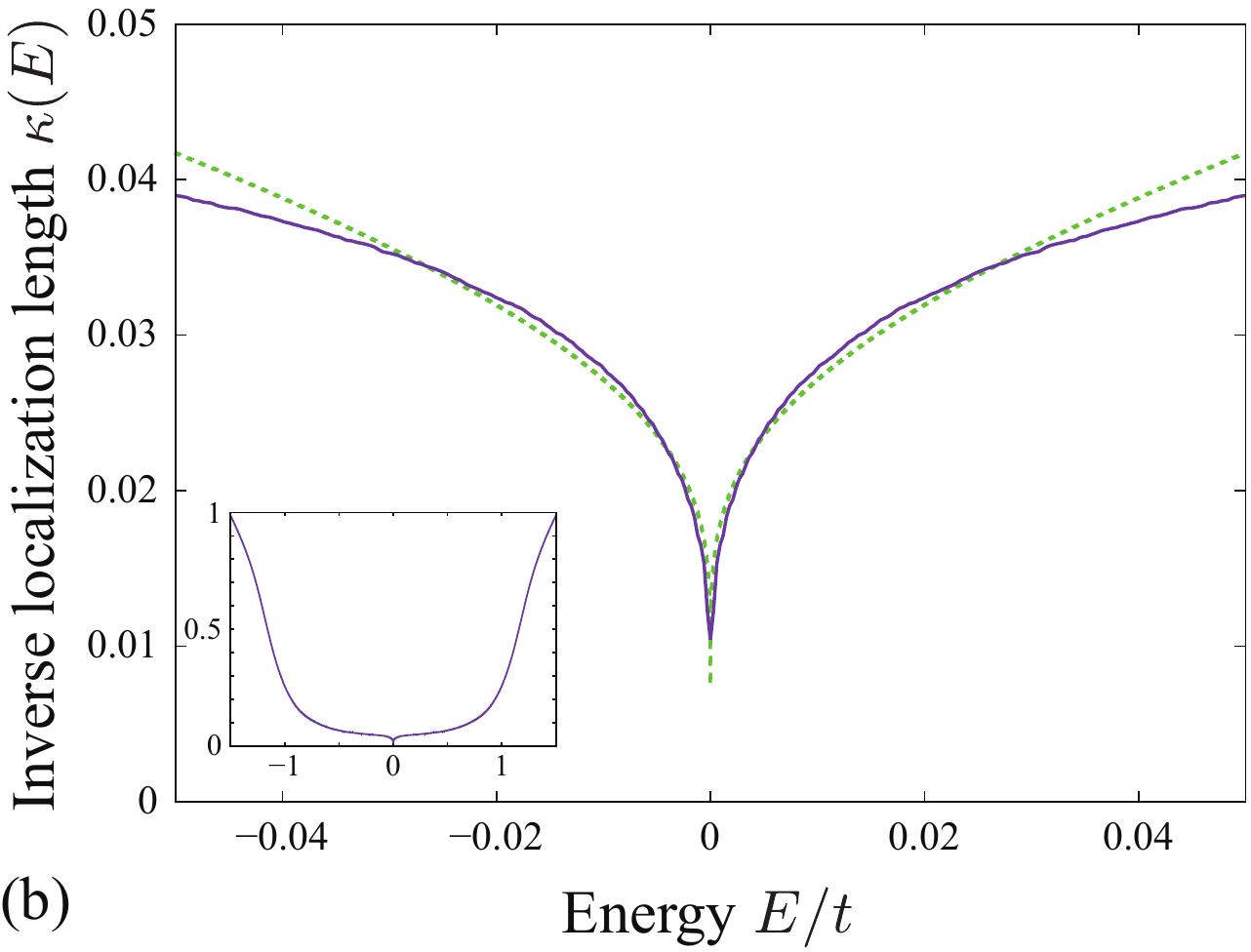}
\caption{The central part of (a) the density of states and (b) the inverse localization length computed from the Chebyshev-polynomial expansion up to the 10 000th order of the random-hopping chain~\eqref{eq200} of length $L=1001$ under open boundary conditions.
The peak height of the density of states in (a) at $E=0$ is $5.76785$ in this particular calculation.
The broken (green) curves in (a) and (b), respectively, follow Eqs.~\eqref{eq201} and~\eqref{eq202} with proportionality constants $3$ and $4$ to guide the eye.
The inset in each panel shows the result in the entire range.
Here we averaged over a set of 1000 random samples.
In both panels, we have removed the normalization of the spectrum into the region $[-1,1]$ and use the original energy scale with $t=1$.}
\label{fig3}
\end{figure}

Notice that the result of our expansion of the localization length $\kappa(E)$ in Fig.~\ref{fig1}~(b) is much smoother than that of the density of states $\rho(E)$ in Fig.~\ref{fig1}~(a).
This is presumably because, as we mentioned at the end of Section~\ref{sec2.2}, higher-order polynomials contribute less in the expansion of $\kappa(E)$ than in the expansion of $\rho(E)$.
We show in Fig.~\ref{fig2} the modulus of the factor $\frac{1}{L}\tr T_{n}(H)$ in the expansion coefficients.
This quantity decays at what appears to be an exponential rate throughout the first 20 to 30 terms (Fig.~\ref{fig2}~(a)), after which it fluctuates around zero. The amplitude of the fluctuation decreases as the square root of the number of samples (Fig.~\ref{fig2}~(b)), which implies that the fluctuation is statistical rather than systematic;
similar behavior of this quantity is observed for the random-hopping model mentioned below.
This is presumably because the Lyapunov exponent is self-averaging~\cite{GLP88,Crisanti93}; as we can see in Eq.~\eqref{eq211}, it is the average over many random terms.

The above observation means that the numerical evaluation of the density of states~\eqref{eq140} requires quite many terms, while that of Eq.~\eqref{eq210} for the inverse localization length can be quite stable; the numerical error due to truncating the series at the 1000th term in the evaluation of the inverse localization length would result in an error less than $10^{-5}$ for only one sample, because the expansion coefficient is divided by the number of the order, and would decrease further as we increase the sample number, whereas that of the density of states would be of order $10^{-2}$, which is indeed the order of fluctuation of the plot in Fig.~\ref{fig1}~(a).

One might alternatively argue that the zigzag features in Fig.~\ref{fig1}~(a) were due to the truncation of the expansion rather than due to the delta peaks of the individual eigenvalues.
In Fig.~\ref{fig1}~(a), we have more than  $10^6$ eigenvalues ($1000$ samples of $1001$ eigenvalues) rather uniformly distributed in the range around $[-1.5,1.5]$, which implies that the average level spacing is about $3\times 10^{-6}$.
On the other hand, the resolution due to the truncation of the expansion is of order $4\times 10^{-3}$, which is too coarse to see the delta peaks of individual eigenvalues.

In contrast, such truncation errors are virtually invisible in Fig.~\ref{fig1}~(b).
This demonstrates the high potential of our Chebyshev-polynomial expansion of the localization length.

Incidentally, we superimpose on Fig.~\ref{fig1}~(b) a numerical estimate of $\kappa(E)$ by an independent method given in Refs.~\cite{HN96,HN97,Hatano98}, where a non-Hermitian extension of the random tight-binding model, also known as the Hatano-Nelson (HN) model, was introduced by making the amplitude of the hopping to the right different from the one to the left:
\begin{align}\label{eq210-1}
H=-\frac{t}{2}\sum_{x=1}^{L}\left(e^g |x+1\rangle\langle x|+e^{-g}|x\rangle\langle x+1|\right)
+\sum_{x=1}^L V_x|x\rangle\langle x|,
\end{align}
where $g$ is a real constant with periodic boundary conditions being assumed.
It was shown in Refs.~\cite{HN96,HN97,Hatano98} that an eigenstate for $g=0$ with the inverse localization length $\kappa$ is delocalized upon increasing the asymmetric parameter $g$ up to $g=\kappa$ and at the same time the corresponding eigenvalue, unchanged (up to small exponential corrections which vanish in the thermodynamic limit of large systems) for $g<\kappa$, gets off the real axis into the complex energy plane.
We can thus obtain an estimation of $\kappa(E)$ of the Hermitian random model, which is superimposed on Fig.~\ref{fig1}~(b), by monitoring the movement of the eigenvalues (\textit{not} the eigenvectors) that the change of the real parameter $g$ gives rise to;
\textit{e.g.}\ in Fig.~\ref{fig1}~(c), the states on the bifurcating endpoints of the bubble of the spectrum for $g=0.5$ would have inverse localization $\kappa=0.5$ for $g=0$.
The result from the model~\eqref{eq210-1} is indeed consistent with the present computation of $\kappa(E)$.

Next, we present the results for the random-hopping model
\begin{align}\label{eq200}
H=-\frac{1}{2}\sum_{x=1}^{L-1}t_x\left(|x+1\rangle\langle x|+|x\rangle\langle x+1|\right);
\end{align}
see Fig.~\ref{fig3}.
Here we sampled the hopping element at each link $t_x$ from a uniform distribution on $[-t-\Delta,-t+\Delta]$ with $\Delta/t=0.5$.
In both Fig.~\ref{fig3}~(a) and~(b), we averaged over a  set of 1000 random samples of length $L=1001$ subjected to open boundary conditions, and terminated the Chebyshev-polynomial expansion at the 10 000th order.

Note the sharp peak of the density of states and a dip of the inverse localization length at $E=0$.
(These are indeed the reason why we terminated the expansion at the high order.) 
It is widely accepted~\cite{Dyson53,Theodorou76,Eggarter78,Ziman82,Brouwer02} that the random-hopping chain~\eqref{eq200} has a diverging density of states around $E=0$ and the corresponding singularity of the localization length:
\begin{align}\label{eq201}
\rho(E)&\sim \left|E(\ln E^2)^3\right|^{-1},
\\\label{eq202}
\kappa(E)&\sim \left|\ln E^2\right|^{-1}.
\end{align}
These are indeed consistent with our data in Fig.~\ref{fig3}.
Since the eigenvalue spectrum of the random-hopping chain~\eqref{eq200} is  symmetric under $E\rightarrow -E$, chains of odd length has an eigenstate with the zero eigenvalue, namely a zero mode.

\begin{figure}
\includegraphics[width=0.4\textwidth]{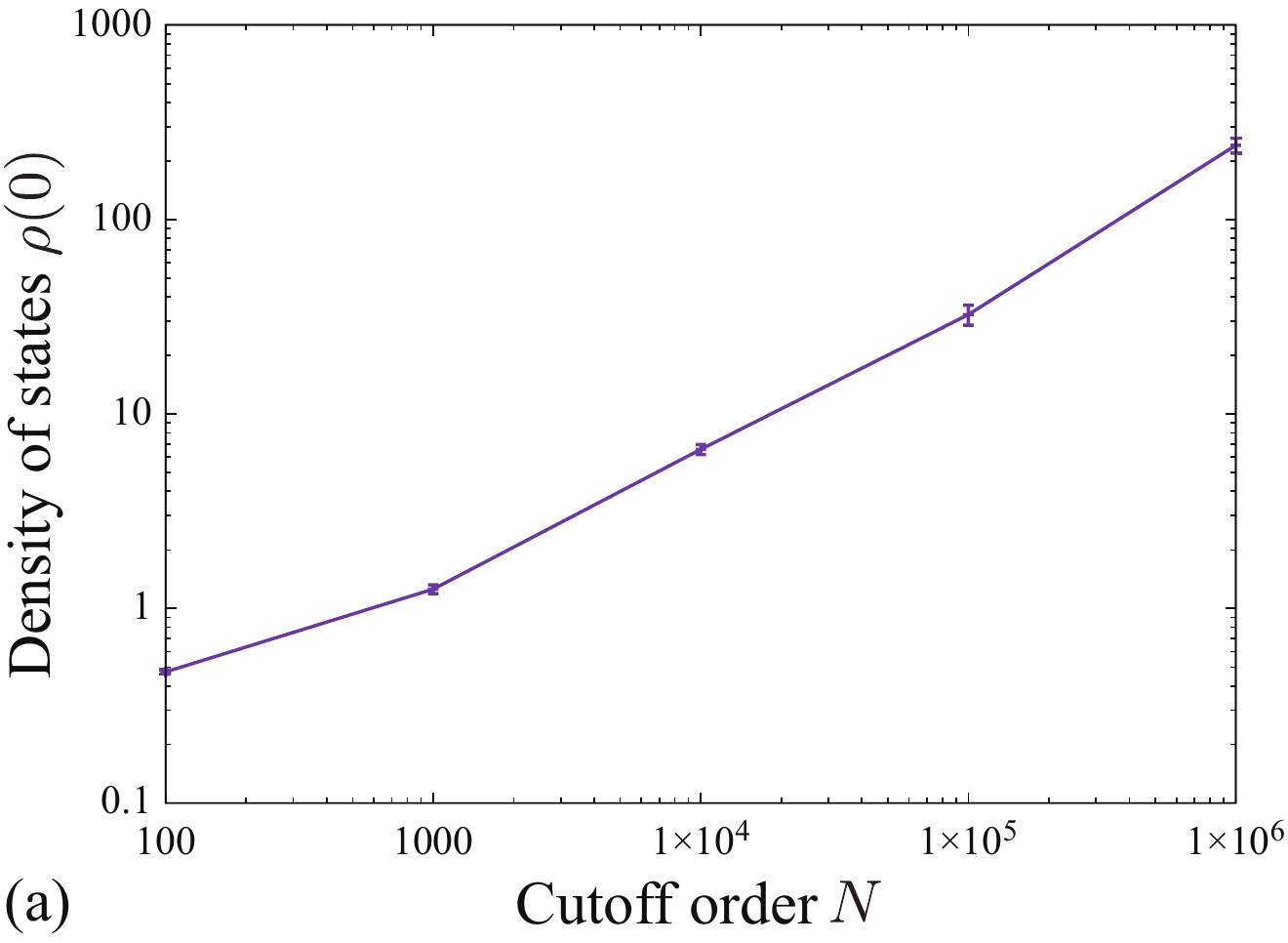}
\\
\vspace{\baselineskip}
\includegraphics[width=0.4\textwidth]{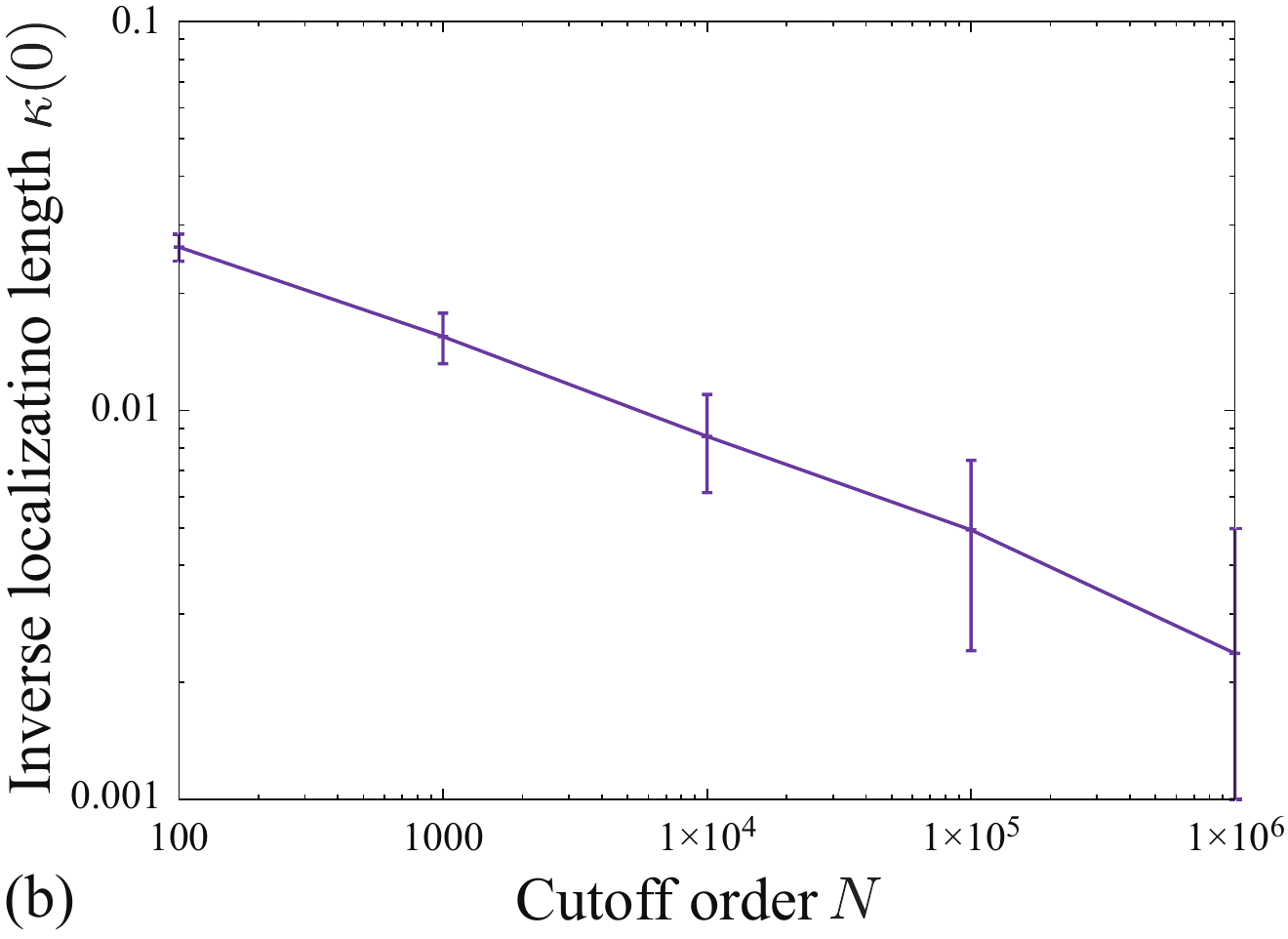}
\caption{(a) The increase of $\rho(0)$ in Eq.~\eqref{eq230} and (b) the decrease of $\kappa(0)$ in Eq.~\eqref{eq240} as we raise the cutoff $N$ of the series
for a random-hopping chain~\eqref{eq200} of length $L=1001$.
We estimated the average and the error from $24$ random samples.}
\label{fig4}
\end{figure}

Let us take a look at the zero mode  from the point of view of the Chebyshev-polynomial expansion.
At $E=0$, the expansions~\eqref{eq140} and~\eqref{eq210} reduce to
\begin{align}\label{eq230}
\rho(0)&=\frac{1}{\pi}\left[1+2\sum_{m=1}^\infty(-1)^m\frac{1}{L}\tr T_{2m}(H)\right],
\\\label{eq240}
\kappa(0)&=-2\sum_{m=1}^\infty(-1)^m\frac{\frac{1}{L}\tr T_{2m}(H)}{2m}-\ln2|\tau|,
\end{align}
where we used $T_{2m}(0) = (-1)^m$ and took advantage of the fact that only the even-order terms contribute because the spectrum of this model is symmetric with respect to $E=0$.
We have noticed in our numerical data that the factor $\tr T_{2m}(H)$ almost alternates in sign with respect to $m$, which makes the series non-alternating when combined with the factor $(-1)^m$.
Because of this behavior, the estimates of $\rho(0)$ and $\kappa(0)$ change monotonically as we increase the cutoff order $N$ of the polynomial.
Figure~\ref{fig4} shows the cutoff-dependence of the two quantities $\rho(0)$ and $\kappa(0)$. 
The former increases and the latter decreases seemingly in power of the cutoff $N$,
which indeed suggests that the density of states diverges and the inverse localization length vanishes at $E=0$.

%

\section{Method for the inverse localization length of non-Hermitian chains}
\label{sec3}

For non-Hermitian hopping matrices, we need a method for computing the density of states $\rho({\rm Re}E, {\rm Im}E)$ and the inverse localization length $\kappa({\rm Re}E, {\rm Im}E)$ in the complex energy plane. These are {\em real} functions of the two real variables ${\rm Re}E, {\rm Im}E$, and are clearly not complex-analytic functions of the complex variable $E$. 
We would therefore need to expand them in double series of orthogonal polynomials,
\begin{align}
\sum_{m,n}c_{m,n}T_m(\re E)T_n(\im E),
\end{align}
for which, however, there are no equivalents of the expansions~\eqref{eq140} and~\eqref{eq210} available.
This is because the non-Hermitian Hamiltonian in question is typically not a normal matrix, \textit{i.e.}\ it does not commute with its adjoint $H^\dag$. Consequently, these two matrices cannot be diagonalized simultaneously.
Therefore, $\sum_\nu (\re E_\nu)^k$ is not simply $\tr [(H+H^\dag)/2]^k$, unlike in Eq.~\eqref{eq130}, and similarly for powers of the imaginary part or products thereof.



\subsection{Method of Hermitization}
\label{sec3.1}

In order to overcome this difficulty, we employ the method of Hermitization invented in~\cite{Feinberg97} (see also~\footnote{Variants of this idea were presented independently in Refs.~\cite{JNPWZ97,JNPZ97,Efetov97a,Efetov97b,Chalker97,Chalker00}. (The content of the present subsection is applicable to systems in any dimensions.) Here we follow the conventions of Ref.~\cite{Feinberg97}. In particular, the dispersion integrals~\eqref{eq230-1},~\eqref{dispersion} and~\eqref{eq280}, expressing $\rho(z,z^*)$ in terms of the eigenvalue density of the Hermitized matrix $\H(z,z^\ast)$, which are essential to our derivations, are unique to Ref.~\cite{Feinberg97}.}). Given an $L\times L$ non-Hermitian Hamiltonian matrix $H$, consider the $2L\times 2L$ `Hermitized' matrix  
\begin{align}\label{eq300}
\H(z,z^\ast)=\begin{pmatrix}
0 & H-z \\
H^\dag - z^\ast & 0
\end{pmatrix}.
\end{align}
The density of states $\rho(z,z^\ast)$ of the non-Hermitian matrix $H$ in terms of the variables
\begin{align}
z=\re E+i\im E
\qquad\mbox{and}\qquad
z^\ast=\re E-i\im E
\end{align}
is given by~\cite{Feinberg97}
\begin{align}\label{eq230-1}
\rho(z,z^\ast)=-\frac{4}{\pi}\int_0^\infty\partial\partial^\ast\frac{\Omega(\mu;z,z^\ast)}{\mu} d\mu,
\end{align}
where
\begin{align}
\partial &= {\partial\over\partial z} 
= \frac{1}{2}\left({\partial\over\partial \re E} - i{\partial\over\partial \im E} \right),
\\
\partial^\ast&={\partial\over\partial z^\ast }
=  \frac{1}{2}\left({\partial\over\partial \re E} + i{\partial\over\partial \im E} \right),
\end{align}
and $\Omega(\mu;z,z^\ast)$ is the {\em integrated} density of states of the Hermitized matrix $\H(z,z^\ast)$.
In other words, 
\begin{align}
\Omega(\mu;z,z^\ast)=\int_{-\infty}^\mu \omega(\mu';z,z^\ast)d\mu',
\end{align}
where $\omega(\mu;z,z^\ast)$ is the density of states of $\H(z,z^\ast)$, supported along the real-$\mu$ axis:
\begin{align}
\omega(\mu;z,z^\ast)=\frac{1}{2L}\sum_{\nu=1}^{2L}\delta(\mu-\mu_\nu(z,z^\ast))\,.
\end{align}
Here $\mu_\nu(z,z^\ast)$ denote the (unknown) eigenvalues of the Hermitized matrix $\H(z,z^\ast)$.
It turns out  that $\omega(\mu;z,z^\ast)$  is an even function of $\mu$, due to the chiral block structure of  $\H(z,z^\ast)$, which implies that eigenvalues of $\H(z,z^\ast)$ come in pairs $\pm\mu_\nu$. 

For later convenience, let us represent Eq.~\eqref{eq230-1} in terms of $\omega$ instead of $\Omega$.
Partial integration gives
\begin{widetext}
\begin{align}\label{eq260}
\rho(z,z^\ast)=-\frac{4}{\pi}\lim_{M\to\infty\atop\epsilon\to0}\int_\epsilon^M
\frac{\partial\partial^\ast\Omega(\mu;z,z^\ast)}{\mu} d\mu
=-\frac{4}{\pi}\lim_{M\to\infty\atop\epsilon\to0}
\left(
\left[\partial\partial^\ast\Omega(\mu;z,z^\ast)\ln\mu\right]_{\mu=\epsilon}^M
-\int_\epsilon^M 
\partial\partial^\ast\omega(\mu;z,z^\ast)\ln\mu\, d\mu\
\right).
\end{align}
\end{widetext}
For a large $M$, the integrated density of states $\Omega(M;z,z^\ast)$ approaches unity, while for a small $\epsilon$, it approaches $1/2$ (because $\omega(\mu;z,z^\ast)$ is an even function of $\mu$), losing the dependence on $z$ and $z^\ast$ in both limits.
The first term in the parentheses of Eq.~\eqref{eq260} therefore vanishes after the derivative $\partial\partial^\ast$.
We thus have
\begin{align}\label{dispersion}
\rho(z,z^\ast)&=\frac{4}{\pi}\int_0^\infty 
\partial\partial^\ast\omega(\mu;z,z^\ast)\ln\mu\,d\mu,
\end{align}
where we took the limit $M\to\infty$, assuming that the density of states has a compact support, and took the limit $\epsilon\to0$ because the integrand now does not have a singularity there.
We can now exchange the integral and the derivative, arriving at
\begin{align}\label{eq280}
\rho(z,z^\ast)&=\frac{4}{\pi}\partial\partial^\ast\int_0^\infty 
\omega(\mu;z,z^\ast)\ln\mu\,d\mu.
\end{align}

\subsection{Method of Hermitization for the inverse localization length}
\label{sec3.2}

Our strategy is now obvious.
We can obtain the Chebyshev-polynomial expansion of the density of states $\rho(z,z^\ast)$ of  $\H(z,z^\ast)$ by applying the method to the density of states $\omega(\mu;z,z^\ast)$, which is supported along the real $\mu$-axis. In fact, we find that the Chebyshev-polynomial expansion of the inverse localization length $\kappa(z,z^\ast)$ is easier to carry than that of $\rho(z,z^\ast)$. In this section we shall focus on $\kappa(z,z^\ast)$, and return to $\rho(z,z^\ast)$  in Section~\ref{sec4}.

In order to find the Chebyshev-polynomial expansion of the inverse localization length $\kappa(z,z^\ast)$, we first need to express $\kappa(z,z^\ast)$ in terms of the density of states $\omega(\mu;z,z^\ast)$ of the Hermitized matrix.
The formula that we utilize is the generalized Thouless formula derived by Derrida \textit{et al.}~\cite{Derrida00}:
\begin{align}
\label{eq290}
\rho(z,z^\ast)
&=\frac{2}{\pi}\partial\partial^\ast\kappa(z,z^\ast)
\end{align}
for non-Hermitian random chains.
In fact, this formula coincides with Eq.(2.9) in Ref.~\cite{Feinberg97} upon the identification
\begin{align}
\kappa(z,z^\ast) = {1\over 2L} \langle\log\det(z-H)^\dagger(z-H)\rangle,
\end{align}
which means that Eq.~\eqref{eq290} holds also for non-Hermitian matrices $H$ more generic than hopping matrices (\textit{e.g.}, non-sparse, completely full matrices, such as the matrices analyzed in Section~\ref{sec4.2}). For such matrices, however,  $\kappa(z,z^\ast)$ in the expression above does not control the spatial decay (or growth) of any of its eigenvectors, losing its meaning as the inverse localization length. Incidentally, noting that Eq.~\eqref{eq290} is the Poisson equation in two dimensions~\cite{Feinberg97}, we obtain its functional inverse~\cite{Derrida00} as
\begin{align}
\kappa(z,z^\ast)=\frac{1}{2}\iint d\zeta  d\zeta^\ast \rho(\zeta,\zeta^\ast)\ln|z-\zeta|,
\end{align}
which is the two-dimensional analog of the Thouless formula~\eqref{eq150}.
Comparing Eq.~\eqref{eq290} with Eq.~\eqref{eq280}, we have
\begin{align}\label{eq310}
\kappa(z,z^\ast)&=2\int_0^\infty 
\omega(\mu;z,z^\ast)\ln\mu\,d\mu+f(z)+g(z^\ast),
\end{align}
where $f(z)$ and $g(z^\ast)$ are arbitrary functions, which we fix hereafter by looking into the limit of $|z|\to\infty$.

To this end, for the sake of concreteness, with no loss of generality, we focus back on hopping matrices. 
For very large values of $|z|$, we can find the inverse localization length $\kappa(z,z^\ast)$ as follows. 
Suppose that the one-dimensional Hamiltonian matrix is given by
\begin{align}
H&=\sum_{x=-\infty}^\infty
\left(t_{x-1,x}|x\rangle\langle x-1|+t_{x+1,x}|x\rangle\langle x+1|\right.
\nonumber\\
&\phantom{=\sigma_{x=-\infty}^\infty}
\left.+V_x|x\rangle\langle x|\right).
\end{align}
The eigenvalue equation $\langle x|H|\psi\rangle=E\langle x|\psi\rangle$ reads
\begin{align}
E\langle x|\psi\rangle&=\langle x|H|\psi\rangle
\nonumber\\
&=t_{x-1,x}\langle x-1|\psi\rangle+t_{x+1,x}\langle x+1|\psi\rangle+V_x\langle x|\psi\rangle.
\end{align}
We can cast this into the form of the transfer matrix as
\begin{align}
&\begin{pmatrix}
\langle x+1|\psi\rangle \\
\langle x|\psi\rangle
\end{pmatrix}
\nonumber\\
&=\begin{pmatrix}
(E-V_x)/t_{x+1,x} & -t_{x-1,x}/t_{x+1,x} \\
1 & 0
\end{pmatrix}
\begin{pmatrix}
\langle x|\psi\rangle \\
\langle x-1|\psi\rangle
\end{pmatrix}.
\end{align}
For a large value of $|E|$, we can ignore $V_x$ in the upper-left element.
The eigenvalues of the transfer matrix are then approximately given by
\begin{align}
\lambda\left(\lambda-\frac{E}{t_{x+1,x}}\right)+\frac{t_{x-1,x}}{t_{x+1,x}}=0,
\end{align}
which is followed by
\begin{align}
\lambda&=\frac{1}{2}\left(\frac{E}{t_{x+1,x}}\pm\sqrt{\frac{E^2}{{t_{x+1,x}}^2}-4\frac{t_{x-1,x}}{t_{x+1,x}}}\right)
\nonumber\\
&\simeq
\frac{E}{t_{x+1,x}},\frac{t_{x-1,x}}{E}.
\end{align}
For the wave function to be normalizable, we choose the second eigenvalue for large values of $|E|$, having
\begin{align}
\langle L|\psi\rangle\simeq\left(\prod_{x=0}^{L-1}\frac{t_{x-1,x}}{E}\right)\langle 0|\psi\rangle.
\end{align}
Identifying them with
\begin{align}
\left|\langle L|\psi\rangle\right| \simeq e^{-\kappa L}\left| \langle 0|\psi\rangle\right|,
\end{align}
we  conclude that
\begin{align}\label{eq190}
\kappa(E)\simeq \ln |E|-\ln |\tau|,
\end{align}
where $|\tau|$ is the geometric mean of $|t_{x-1,x}|$.
This is the behavior of the left-hand side of Eq.~\eqref{eq310} for large values of $|z| = |E|$.

On the other hand, we can find the right-hand side of Eq.~\eqref{eq310} for large values of $|z|$ as follows.
The Hermitized matrix $\H(z,z^\ast)$ in Eq.~\eqref{eq300} is reduced to
\begin{align}
\H\to\begin{pmatrix}
0 & -z \\
-z^\ast & 0 
\end{pmatrix}
\end{align}
for large values of $|z|$, and hence its eigenvalues degenerate into $\mu_\nu=\pm|z|$, which means
\begin{align}
\omega(\mu;z,z^\ast)=\frac{1}{2}\left(\delta(\mu-|z|)+\delta(\mu+|z|)\right).
\end{align}
Therefore, the first term on the right-hand side of Eq.~\eqref{eq310} is reduced to $\ln|z|$,
which is the same as the first term of the right-hand side of Eq.~\eqref{eq190}.

The remaining terms in Eq.~\eqref{eq310}, $f(z)+g(z^\ast)$, therefore should converge to the constant value $-\ln |\tau|$ for large values of $|z|$.
An analytic function in general attains its maximum in a given domain on the boundary of that domain.
Moreover, a bounded analytic function is necessarily a constant. 
Since both $f(z)$ and $g(z^\ast)$ are bounded, they are constants;
they cannot balance each other, since one is holomorphic and the other is anti-holomorphic. 
In other words, we can fix the arbitrary functions as
\begin{align}
f(z)+g(z^\ast)=-\ln |\tau|
\end{align}
for all $z$ and $z^\ast$.

We therefore arrive at the relation
\begin{align}\label{eq440}
\kappa(z,z^\ast)&=2\int_0^\infty
\omega(\mu;z,z^\ast)\ln\mu\,d\mu-\ln |\tau|.
\end{align}
Comparing this to Eq.~\eqref{eq280}, we note that we have gotten rid of the differentiation $\partial\partial^\ast$ here.
This is the reason why the Chebyshev-polynomial expansion of $\kappa(z,z^\ast)$ is easier than that of $\rho(z,z^\ast)$.

\subsection{Chebyshev-polynomial expansion of the inverse localization length}
\label{sec3.3}

We are now in a position to find the Chebyshev-polynomial expansion of the inverse localization length $\kappa(z,z^\ast)$ by applying the method to the density of states $\omega(\mu;z,z^\ast)$ on the real axis of $\mu$.
Assume that the density of states of the Hermitized matrix is expanded in the form
\begin{align}\label{eq240-1}
\omega(\mu;z,z^\ast)=\frac{1}{\sqrt{1-\mu^2}}\sum_{m=0}^\infty c_{2m}(z,z^\ast)T_{2m}(\mu),
\end{align}
where we used only  even-order terms because $\omega(\mu;z,z^\ast)$ is an even function of $\mu$, thanks to the chiral block structure of the Hermitized matrix $\H$.

We repeat here the important remark that we can employ the Chebyshev polynomial expansion only after we have properly rescaled the non-Hermitian Hamiltonian $H$ in such a way that the support of the spectrum $\omega(\mu;z,z^\ast)$ of the Hermitized matrix $\H$ is contained in $[-1,1]$.
Let us find the proper rescaling factor.
Suppose that the sum of the moduli of all elements in a row or a column of the non-Hermitian matrix $H$ is bounded by a constant $\Lambda$, which also bounds the spectrum, according to the Gershgorin circle theorem.
Let us then rescale $H$ by a factor $r$, so that the bound is rescaled as $\Lambda/r$.
It is then enough to scan the spectrum over the range $|\re E|\leq \Lambda/r$ and $|\im E|\leq \Lambda/r$, which means that it is enough to investigate the spectrum in the range  $|z|=|z^\ast|\leq\sqrt{2}\Lambda/r$.
Therefore, the sum of the moduli of all elements in a row or a column of the Hermitized matrix $\H(z,z^\ast)$ is bounded by $(1+\sqrt{2})\Lambda/r$ if we scan the spectrum only over the above domain in the complex $z$-plane. The support $[-1,1]$ of the Chebyshev polynomials (along the $\mu$-axis) should contain this bound.
We therefore rescale the non-Hermitian Hamiltonian $H$ with a rescaling factor $r$ that is equal to or greater than $(1+\sqrt{2})\Lambda$.
We assume that the matrix $H$ has been already normalized in this way throughout this subsection.

Let us come back to Eq.~\eqref{eq240-1} and find the expansion coefficients using the orthogonality relation~\eqref{eq30}.
In a manner similar to Eqs.~\eqref{eq110}--\eqref{eq120} we find that the coefficients are given by
\begin{align}\label{eq460}
c_{2m}(z,z^\ast)&=\frac{2}{\pi}\int_{-1}^1 \omega(\mu;z,z^\ast)T_{2m}(\mu)d\mu
\nonumber\\
&=\frac{2}{\pi}\frac{1}{2L}\sum_{\nu=1}^{2L}T_{2m}(\mu_\nu)
=\frac{2}{\pi}\frac{1}{2L}\tr T_{2m}(\H(z,z^\ast))
\end{align}
for $m\geq1$ and
\begin{align}
c_{0}(z,z^\ast)&=\frac{1}{\pi}\int_{-1}^1 \omega(\mu;z,z^\ast)T_0(\mu)d\mu
\nonumber\\
&=\frac{1}{\pi}\int_{-1}^1 \omega(\mu;z,z^\ast)d\mu
=\frac{1}{\pi}.
\end{align}
The trace on the right-hand side of Eq.~\eqref{eq460} is recursively generated from
\begin{align}\label{eq340}
T_{n+1}(\H)=2\H T_n(\H)-T_{n-1}(\H).
\end{align}

Recall that we need only the even-order Chebyshev polynomials of $\H$.
As we show in Appendix~\ref{appB}, the even-order matrix polynomials $T_{2m}(\H)$ have non-vanishing elements only in the $L\times L$ upper-left and lower-right diagonal blocks, whereas the odd-order ones have their non-vanishing elements only on the off-diagonal blocks.
We also prove in Appendix~\ref{appB} that for the even-order polynomials of $\H$, the trace of the upper-left block is equal to the trace of the lower-right block.
We can therefore reduce the matrix size of the recursion relation~\eqref{eq340} from $2L\times 2L$ to $L\times L$:
\begin{align}
T^{(1,1)}_{2m}&=2(H-z)T^{(2,1)}_{2m-1}-T^{(1,1)}_{2m-2},
\\
T^{(2,1)}_{2m+1}&=2(H^\dag-z^\ast)T^{(1,1)}_{2m}-T^{(2,1)}_{2m-1},
\end{align}
where the superscripts $(1,1)$ and $(2,1)$ denote the $L\times L$ upper-left and lower-left blocks, respectively (with obvious similar notation for the remaining blocks).
We can save computer memory storage by using these recursion relations and write
\begin{align}
c_{2m}(z,z^\ast)&=\frac{2}{\pi}\frac{1}{L}\tr T^{(1,1)}_{2m}(\H(z,z^\ast))
\end{align}
instead of Eq.~\eqref{eq460}.

We now plug in the expansion~\eqref{eq240-1} into Eq.~\eqref{eq440} to have
\begin{align}\label{eq360}
\kappa(z,z^\ast)&=-\ln |\tau|
+\frac{2}{\pi}\int_0^1 \frac{\ln\mu}{\sqrt{1-\mu^2}}d\mu
\nonumber\\
&+\frac{4}{\pi}\sum_{m=1}^\infty \frac{1}{L}\tr T^{(1,1)}_{2m}(\H(z,z^\ast))
\nonumber\\
&\times\left( \int_0^1 T_{2m}(\mu)\frac{\ln\mu}{\sqrt{1-\mu^2}}d\mu\right),
\end{align}
where  we reduced the upper limit of the integration range from $\infty$ to $1$ because we rescaled $H$ so that $\omega(\mu;z,z^\ast)$ vanishes beyond unity.
The integral in the second term on the right-hand side of Eq.~\eqref{eq360} is given by the formula
\begin{align}
\int_0^1 \frac{\ln\mu}{\sqrt{1-\mu^2}}d\mu=-\frac{\pi}{2}\ln2.
\end{align}
The other integrals for $m\geq 1$ are given by substituting $E=0$ in Eqs.~\eqref{eq170} and~\eqref{eq180}:
\begin{align}\label{eq370}
\int_0^1 T_{2m}(\mu)\frac{\ln\mu}{\sqrt{1-\mu^2}}d\mu
=-\frac{\pi}{2}\frac{(-1)^m}{2m};
\end{align}
see Appendix~\ref{appC} for an alternative derivation.
We therefore arrive at the expansion of the inverse localization length in the form
\begin{align}\label{eq410}
\kappa(z,z^\ast)&=-\sum_{m=1}^\infty \frac{(-1)^m}{m}\frac{1}{L}\tr T^{(1,1)}_{2m}(\H(z,z^\ast))-\ln (2|\tau|)\,.
\end{align}
We show in Appendix~\ref{appD} that this indeed reduces to Eq.~\eqref{eq210} when $H$ is Hermitian.

\subsection{Demonstration}
\label{sec3.4}

Let us now demonstrate our new algorithm of the Chebyshev-polynomial expansion~\eqref{eq410}.
In Figs.~\ref{fig5} and~\ref{fig7} below, we have removed the normalization of the spectrum and plotted the results in the original energy scale.

We here use a random-sign model, also known as the Feinberg-Zee (FZ) random-hopping model~\cite{FZ99,Holz03,Chandler-Wilde11,Chandler-Wilde12,Chandler-Wilde13,Chandler-Wilde15,Hagger14,Hagger15a,Hagger15b,AHN15}, defined by the Hamiltonian
\begin{align}\label{eq610}
H=\sum_{x=1}^L\left(t_x|x+1\rangle\langle x|+s_x|x\rangle\langle x+1|\right),
\end{align}
where each of the hopping amplitudes  $\{t_x\}$ and $\{s_x\}$ is independently randomly chosen from $\pm 1$ with equal probabilities; 
periodic boundary conditions are assumed.
The spectrum is a fuzzy fractal-like object as is shown in Fig.~\ref{fig5}~(a);
note the exact (deterministic) reflection symmetries with respect to the real and imaginary axes as well as the statistical reflection symmetries with respect to the $45^\circ$ and $135^\circ$ lines~\cite{FZ99,AHN15}.
The deterministic symmetries are easy to understand. Since $H$ in Eq.~\eqref{eq610} is a real matrix, its complex eigenvalues come in complex conjugate pairs $E,E^*$, which means the symmetry of the spectrum against reflections with respect to the real axis. The spectrum of $H$ is also symmetric with respect to reflection through the origin. This symmetry arises from the fact that the diagonal matrix $D$ with alternating $\pm1$ diagonal elements, {\em anticommutes}  with $H$ (subjected to periodic boundary conditions), provided the length $L$ is even. (For open boundary conditions, these matrices anticommute for any $L$.) Thus, eigenvalues of $H$ come in pairs $\pm E$. Combining these two symmetries, we see that complex eigenvalues come in quadruplets $\pm E, \pm E^*$.
\begin{figure*}
\includegraphics[width=0.4\textwidth]{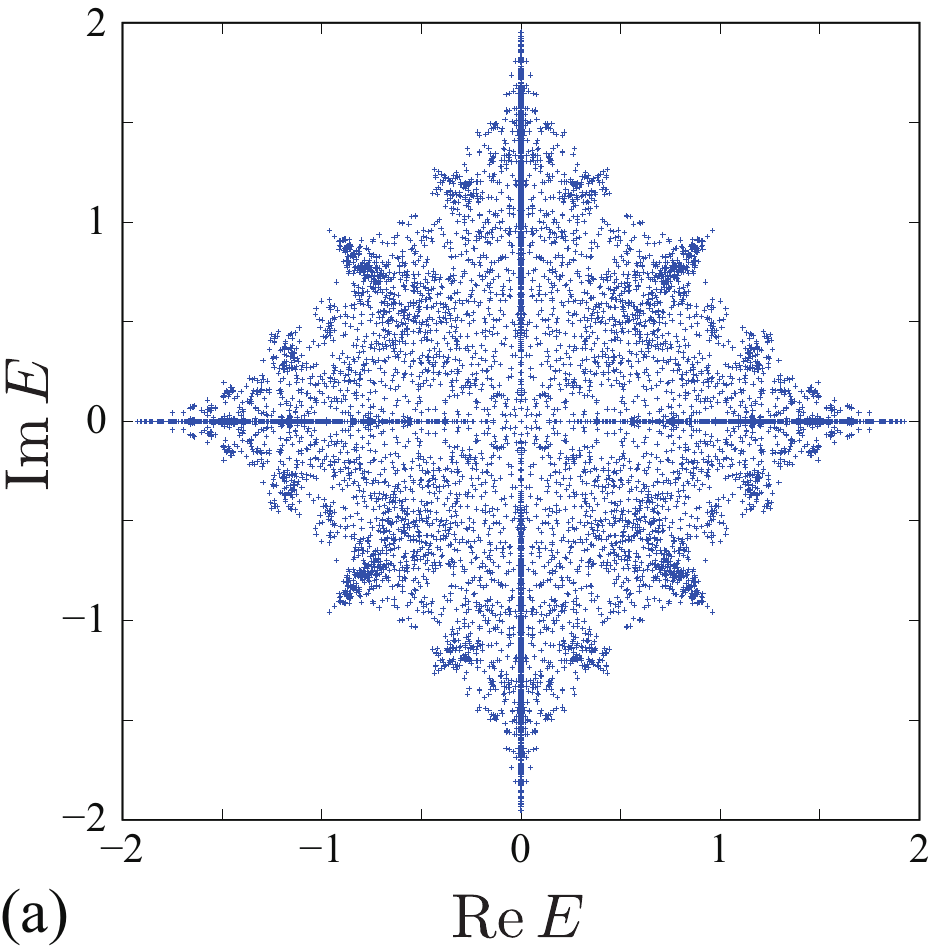}
\hspace{0.06\textwidth}
\includegraphics[width=0.4\textwidth]{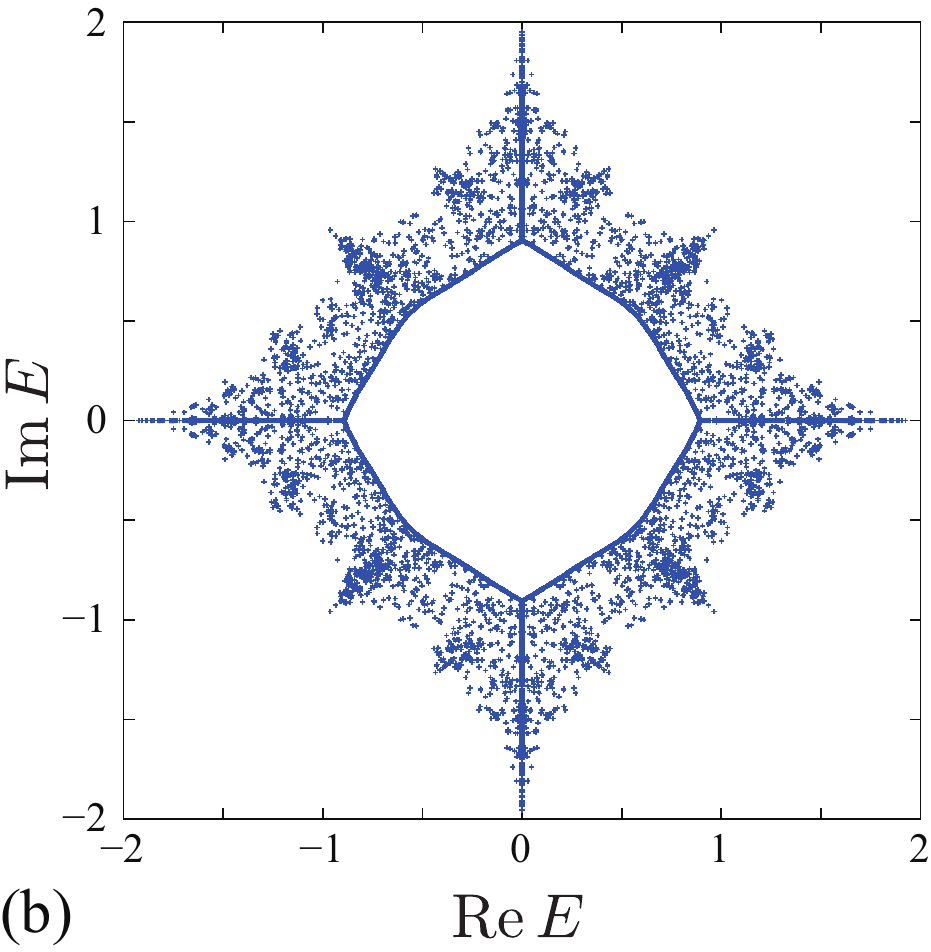}
\vspace{\baselineskip}
\\
\includegraphics[width=0.45\textwidth]{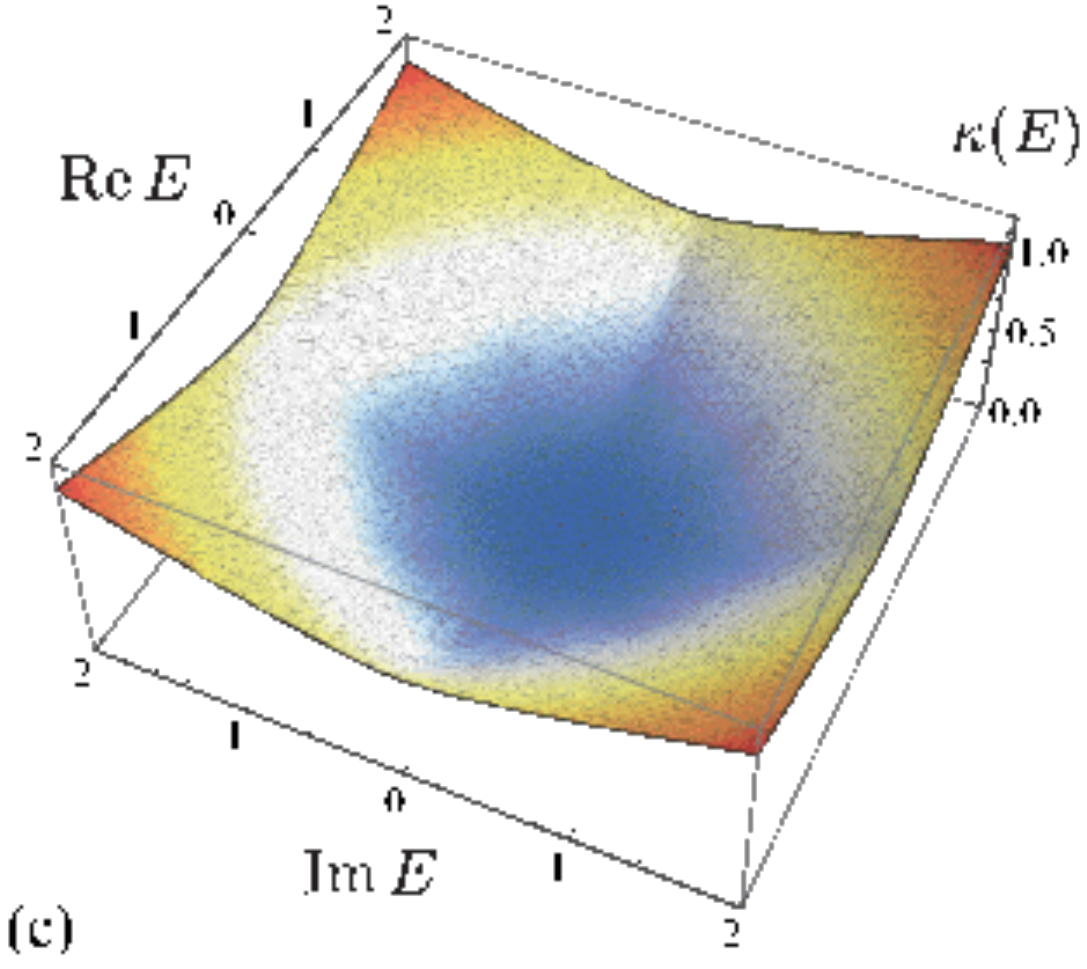}
\hspace{0.03\textwidth}
\includegraphics[width=0.4\textwidth]{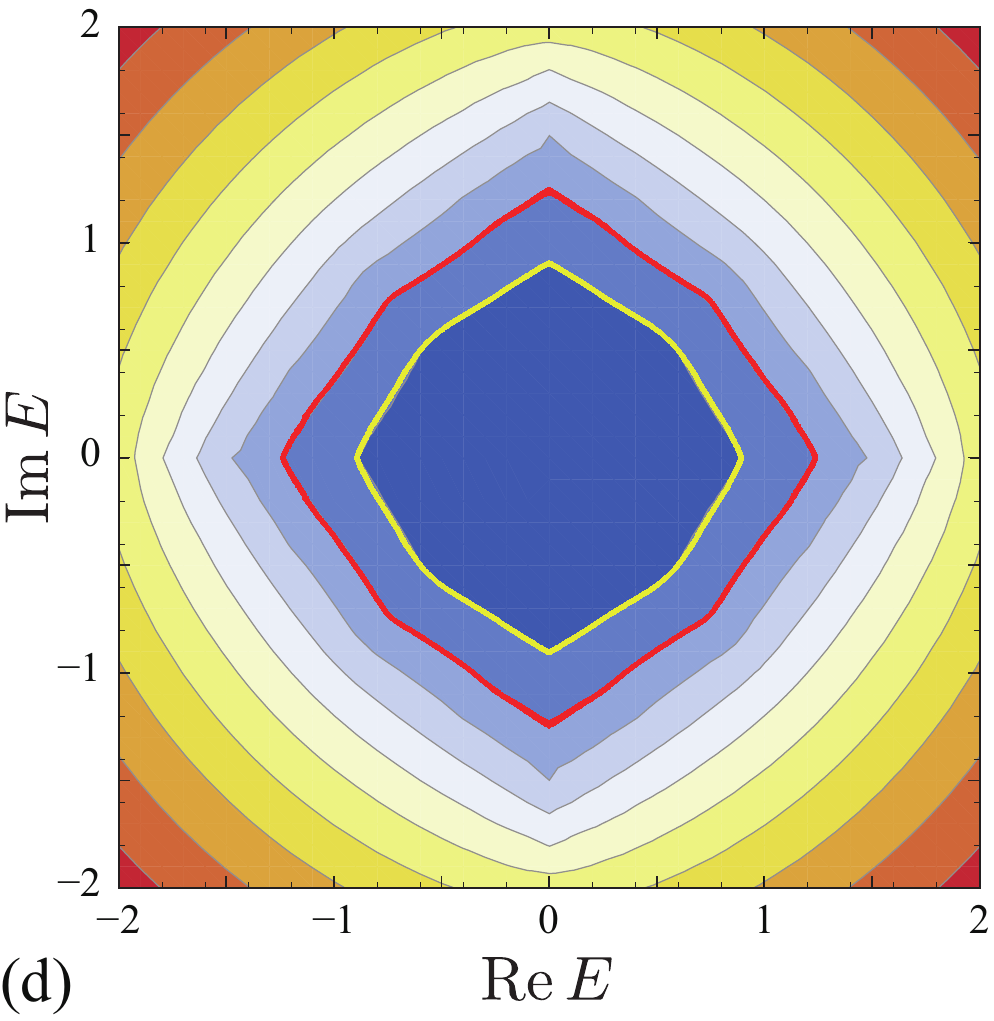}
\caption{Plots of the eigenvalue distributions of (a) the FZ random-hopping model~\eqref{eq610} and (b) its HN-gauged version~\eqref{eq620}, both for chains of length $L=10000$, obtained by direct numerical diagonalization.
(c) A three-dimensional plot and (d) a contour plot of the result of the expansion~\eqref{eq410} up to the 500th order, averaged over 96 random samples of the FZ random-hopping model~\eqref{eq610} of length $L=100$.
The level contours in the panel (d) indicate the data from $\kappa(E)=0.1$ inside to $1.0$ outside in increments of 0.1.
The thick yellow and red curves superimposed on the panel (d) indicate the rims of the hole in the spectrum of the HN-gauged model with an asymmetric field $g=0.1$ (the spectrum in the panel (b)) and $g=0.2$ added~\cite{AHN15}, respectively.
Here we have removed the normalization of the spectrum and plot the results in the original energy scale.}
\label{fig5}
\end{figure*}

The statistical symmetry comes from the fact that the statistics of the matrix does not change after multiplying it by a factor $i$ (or $-i$).
This is so because, as explained in Sec.~IV of Ref.~\cite{FZ99} and Sec.~II of Ref.~\cite{AHN15}, the spectrum of $H$ in Eq.~\eqref{eq610} depends only on products of pairs of opposing off-diagonal, hopping matrix elements $R_x=s_xt_x$.
In our model, $\{R_x|1\leq x\leq L\}$ are statistically independent of each other and each takes on values $\pm 1$ with equal probabilities. 
Multiplying $H$ by a factor $i$ is equivalent to choosing an equally probable element on the ensemble with all $\{R_x\}$ reversed in sign, and hence, on average, does not change the spectrum.
The multiplication, on the other hand, rotates the entire spectrum by $90^\circ$ on the complex energy plane.
Thus, on average, the spectrum of $H$ should be symmetric against rotation by $\pm 90^\circ$.

The result of our expansion~\eqref{eq410} for the inverse localization length is shown in Fig.~\ref{fig5}~(c) and~(d).
This is basically consistent with the result in Fig.~9~(a) of Ref.~\cite{AHN15}, where the inverse localization length of the FZ random-hopping model~\eqref{eq610} was estimated by means of the transfer-matrix approach as the average of the logarithm of the Ricatti variable $\left.\langle x+1|\psi\rangle\middle/\langle x|\psi\rangle\right.$.

\begin{figure}
\includegraphics[width=0.42\textwidth]{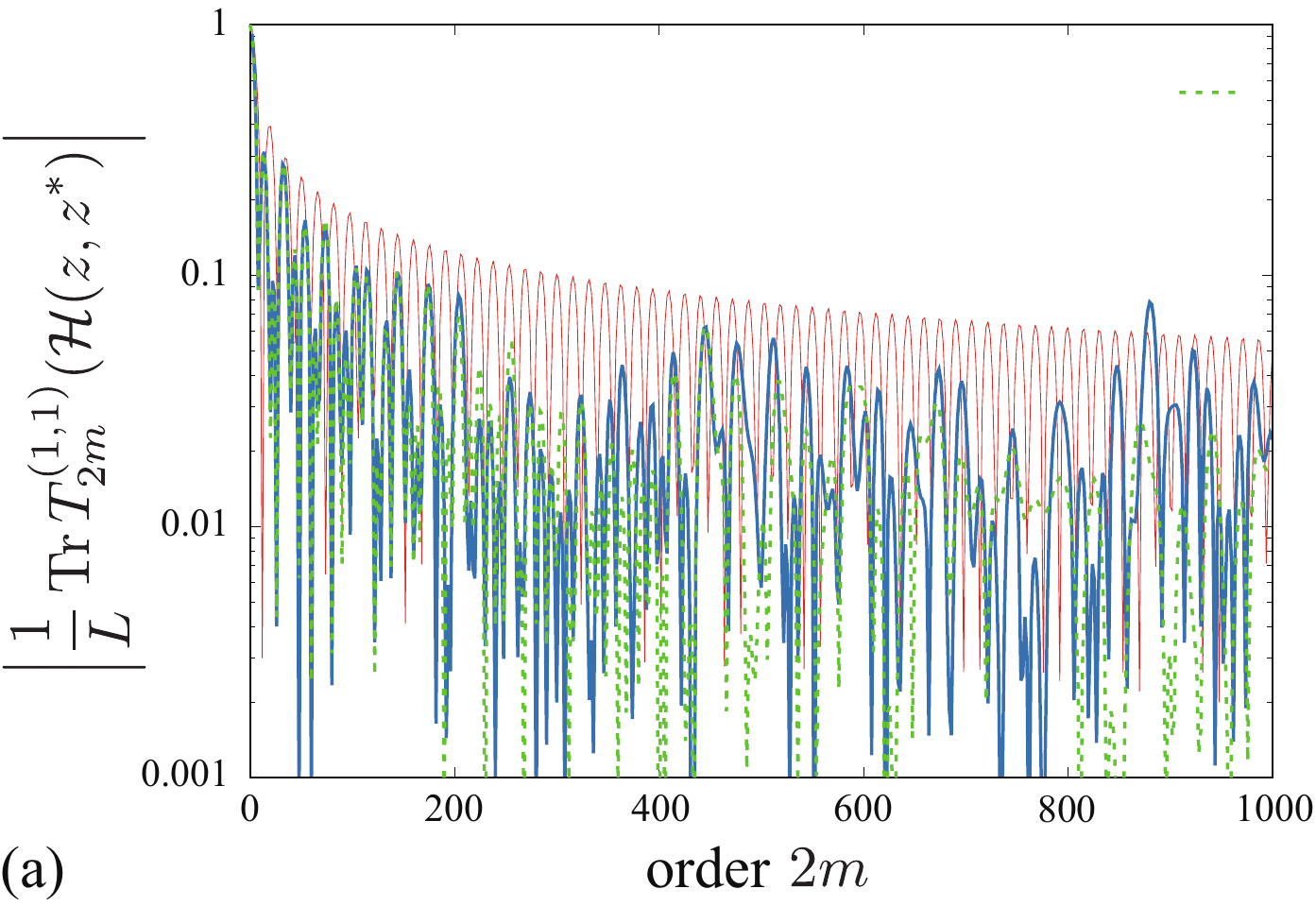}
\\
\vspace{\baselineskip}
\includegraphics[width=0.42\textwidth]{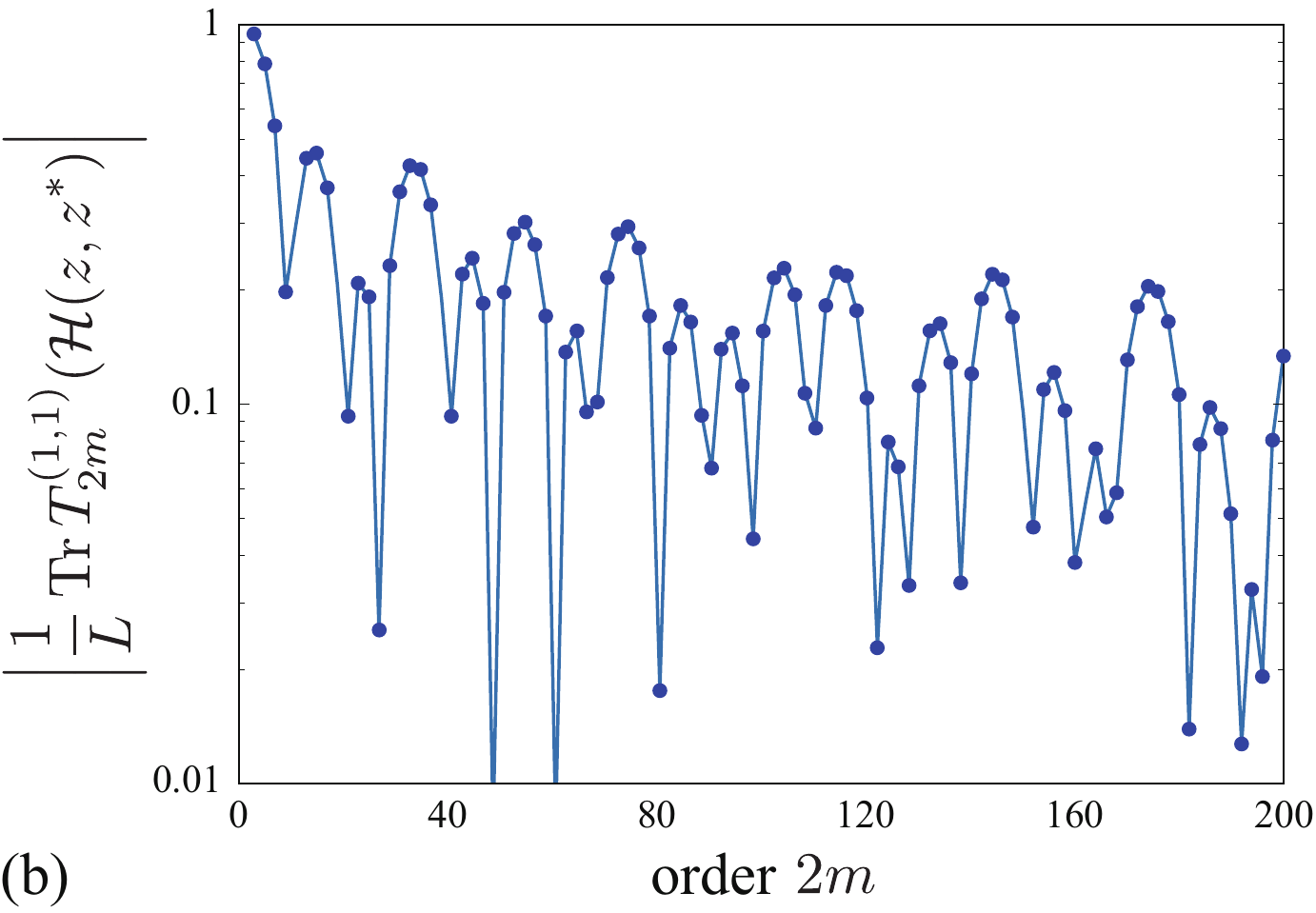}
\caption{Semi-logarithmic plots of the modulus of $\frac{1}{L}\tr T^{(1,1)}_{2m}(\H(z,z^\ast))$ (a) at $z=1+i$ (thicker curve) and at $z=0$ (thinner curve) to the 1000th order, and 
(b) zooming into the left part of the upper figure for $z=1+i$, up to the 200th order.
We here used only one random sample of the random-sign chain~\eqref{eq610} of length $L=1000$, except that  in (a) we also plotted the average over 10 008 samples (broken green line).}
\label{fig6}
\end{figure}

Note the smoothness of the result in Fig.~\ref{fig5}~(c).
We attribute it again to the factor $1/m$ in the expansion~\eqref{eq410}, as we did at the end of Section~\ref{sec2.2} for Hermitian models.
We display in Fig.~\ref{fig6} the modulus of the factor $\frac{1}{L}\tr T^{(1,1)}_{2m}(\H(z,z^\ast))$ in the expansion coefficients.
At $z=1+i$, the factor decays almost exponentially up to the 200th order but then the modulus fluctuates around 0.01 with a possible slight decay.
This behavior is qualitatively similar for almost any value of $z$, except for $z=0$, where this coefficient decays regularly, as shown in Fig.~\ref{fig6}~(b).
The cutoff error is suppressed thanks to the factor $1/m$ in the expansion coefficient.
Incidentally, as can be seen from Fig.~\ref{fig6}~(a), these fluctuations do not seem to depend on the number of samples, which implies that it is not statistical, unlike the case in Fig.~\ref{fig2}~(b); 
the convergence to the self-averaged value of the Lyapunov exponent may not be uniform because of the additional $z$ dependence.

While many studies~\cite{FZ99,Holz03,Chandler-Wilde11,Chandler-Wilde12,Chandler-Wilde13,Chandler-Wilde15,Hagger14,Hagger15a,Hagger15b} had focused on the fractal-like spectrum of the FZ random-hopping model~\eqref{eq610}, the study in Ref.~\cite{AHN15} found two new features of the model's inverse localization length, which we here reproduce by means of our Chebyshev-polynomial expansion~\eqref{eq410}.
First, the inverse localization length was found numerically to behave interestingly around $E=0$~\cite{AHN15}.
We can prove that the inverse localization length vanishes at $E=0$~\cite{AHN15} just as in the random-hopping model~\eqref{eq200}, but the behavior around $E=0$~\cite{Dyson53,Theodorou76,Eggarter78,Ziman82,Brouwer02} seems to be very different from Eq.~\eqref{eq202}.
Numerical data of the Chebyshev-polynomial expansion in Fig.~\ref{fig7} seems to be consistent with small-energy behavior $\kappa(E,E^\ast)\sim |E|^2 f(\arg E)$ for some function $f$ of the argument of the complex energy $E$.
\begin{figure}
\includegraphics[width=0.4\textwidth]{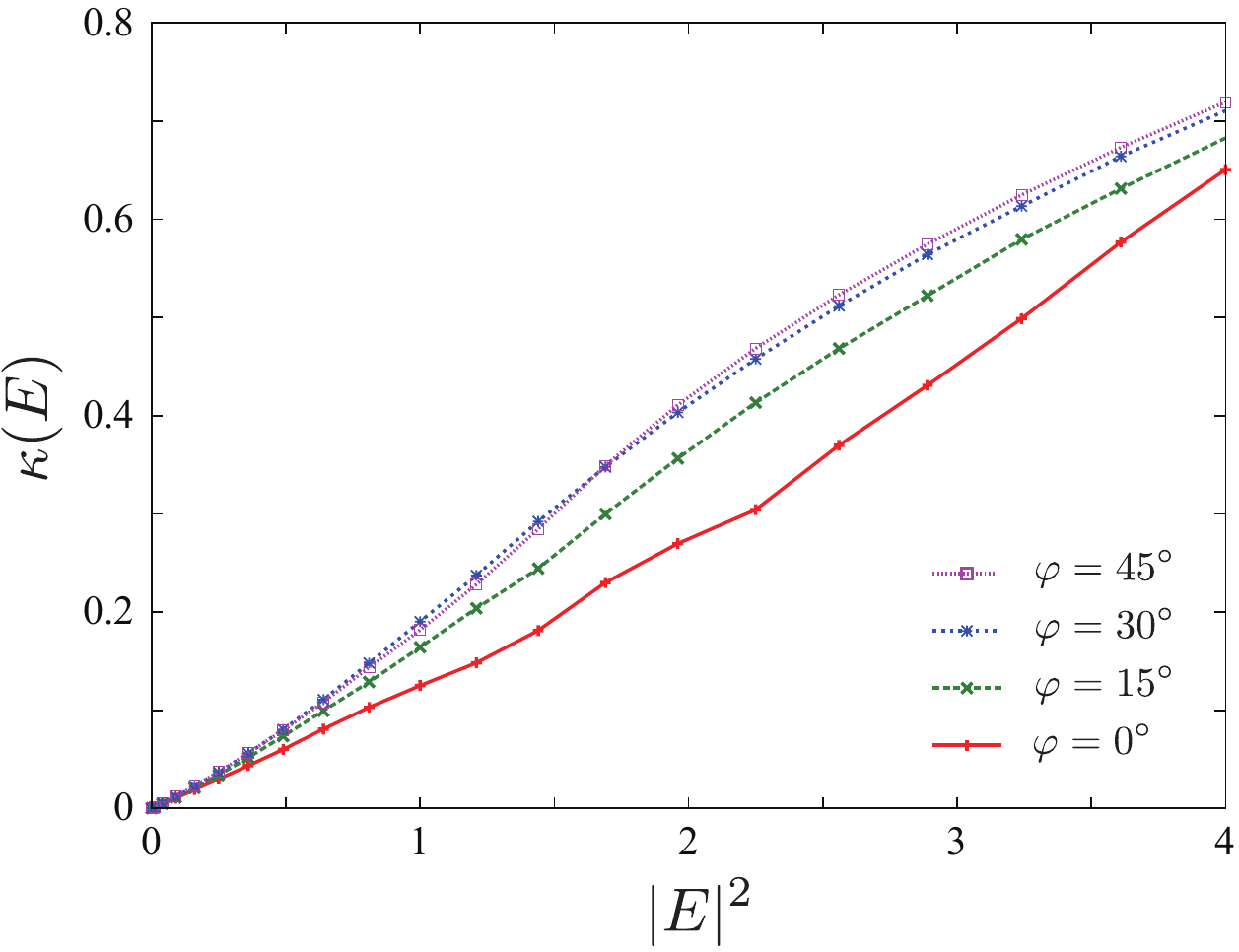}
\caption{The inverse localization length $\kappa(E)$ at $E=|E|e^{i\varphi}$ according to the expansion~\eqref{eq410} up to the 500 order of the random-sign chain~\eqref{eq610} of length $L=500$
for four values of $\varphi$ averaged over 96 random samples.
Note that the horizontal axis indicates $|E|^2$.
Here we have removed the normalization of the spectrum and plotted the results in the original energy scale.}
\label{fig7}
\end{figure}
The linearity with respect to $|E|^2$ seen in Fig.~\ref{fig7} implies a non-singular density of states around $E=0$ according to Derrida's formula~\eqref{eq290}, which is indeed consistent with the results in Ref.~\cite{AHN15}, where the density of states is shown to be vanishing around $E=0$ in a non-singular way.

The other feature found in Ref.~\cite{AHN15} for the FZ random-hopping model~\eqref{eq610} concerns the ``HN-gauged" FZ random-hopping model introduced in the spirit of the model~\eqref{eq210-1}:
\begin{align}\label{eq620}
H=\sum_{x=1}^L\left(e^g t_x|x+1\rangle\langle x|+e^{-g}s_x|x\rangle\langle x+1|\right),
\end{align}
where $g$ is again a real parameter, and periodic boundary conditions are assumed.
As we increase $g$, a hole opens up in the fractal-like spectrum as exemplified in Fig.~\ref{fig5}~(b) for $g=0.1$.
It was conjectured~\cite{AHN15} that the eigenstates that reside on the rim of the hole for $g>0$ had the inverse localization length $\kappa=g$ for $g=0$, as was the case for the model~\eqref{eq210-1}.
As an example, the rims of the hole in the spectrum of the HN-gauged model~\eqref{eq620} for $g=0.1$ (the one in Fig.~\ref{fig5}~(b)) and $g=0.2$ are superimposed on the contour plot of $\kappa(E)$ of the model~\eqref{eq610} in Fig.~\ref{fig5}~(d).
The rims of the hole fall nicely on to the contours of $\kappa(E)=0.1$ and $0.2$, which indeed supports the conjecture.

\section{Method for the density of states of non-Hermitian matrices}
\label{sec4}

\subsection{Chebyshev-polynomial expansion of the density of states}
\label{sec4.1}

We finally give the algorithm for the density of states of non-Hermitian matrices. All we have to do is to plug the expansion~\eqref{eq410} for $\kappa(z,z^\ast)$ into Eq.~\eqref{eq290}. 
Note that the content of the present section is applicable to systems in any dimensions;
in particular, we will demonstrate below how the algorithm works for an example of full random matrices.
Thus, we obtain
\begin{align}\label{eq400}
\rho(z,z^\ast)&=-\frac{4}{\pi}\sum_{m=1}^{\infty}\frac{(-1)^m}{2m}
\frac{1}{L}\tr \partial\partial^\ast T_{2m}^{(1,1)}(\H(z,z^\ast))
\\\label{eq400-1}
&=-\frac{4}{\pi}\sum_{m=1}^{\infty}\frac{(-1)^m}{2m}
\frac{1}{2L}\tr \partial\partial^\ast T_{2m}(\H(z,z^\ast))\,,
\end{align}
where in the last equation we restored the full $2L\times 2L$ matrix (hence the extra 1/2 factor). We can generate the factor $\frac{1}{2L}\tr T_{2m}(\partial\partial^\ast\H(z,z^\ast))$ in Eq.~\eqref{eq400-1} recursively as follows.
By differentiating Eq.~\eqref{eq340}, we have
\begin{align}
\partial T_{n+1}(\H)&=
-2
\begin{pmatrix}
0 & I \\
0 & 0
\end{pmatrix}
T_{n}(\H)
\nonumber\\
&+2\H\partial T_{n}(\H)-\partial T_{n-1}(\H),
\\
\partial^\ast T_{n+1}(\H)&=
-2
\begin{pmatrix}
0 & 0 \\
I & 0
\end{pmatrix}
T_{n}(\H)
\nonumber\\
&+2\H\partial^\ast T_{n}(\H)-\partial^\ast T_{n-1}(\H),
\\
\partial\partial^\ast T_{n+1}(\H)&=
-2
\begin{pmatrix}
0 & 0 \\
I & 0
\end{pmatrix}
\partial T_{n}(\H)
-2
\begin{pmatrix}
0 & I \\
0 & 0
\end{pmatrix}
\partial^\ast T_{n}(\H)
\nonumber\\
&+2\H\partial\partial^\ast T_{n}(\H)-\partial\partial^\ast T_{n-1}(\H).
\end{align}
We can thereby generate the series of $\partial T_{n}(\H)$ and the series of $\partial^\ast T_{n}(\H)$ with the help of the series $T_n(\H)$, and finally the series of $\partial\partial^\ast T_{n}(\H)$ with the help of the preceding two series.
We therefore need four matrix multiplications to generate one more element in the series of $\partial\partial^\ast T_{n}(\H)$.

\subsection{Demonstration}
\label{sec4.2}

\begin{figure}
\includegraphics[width=0.42\textwidth]{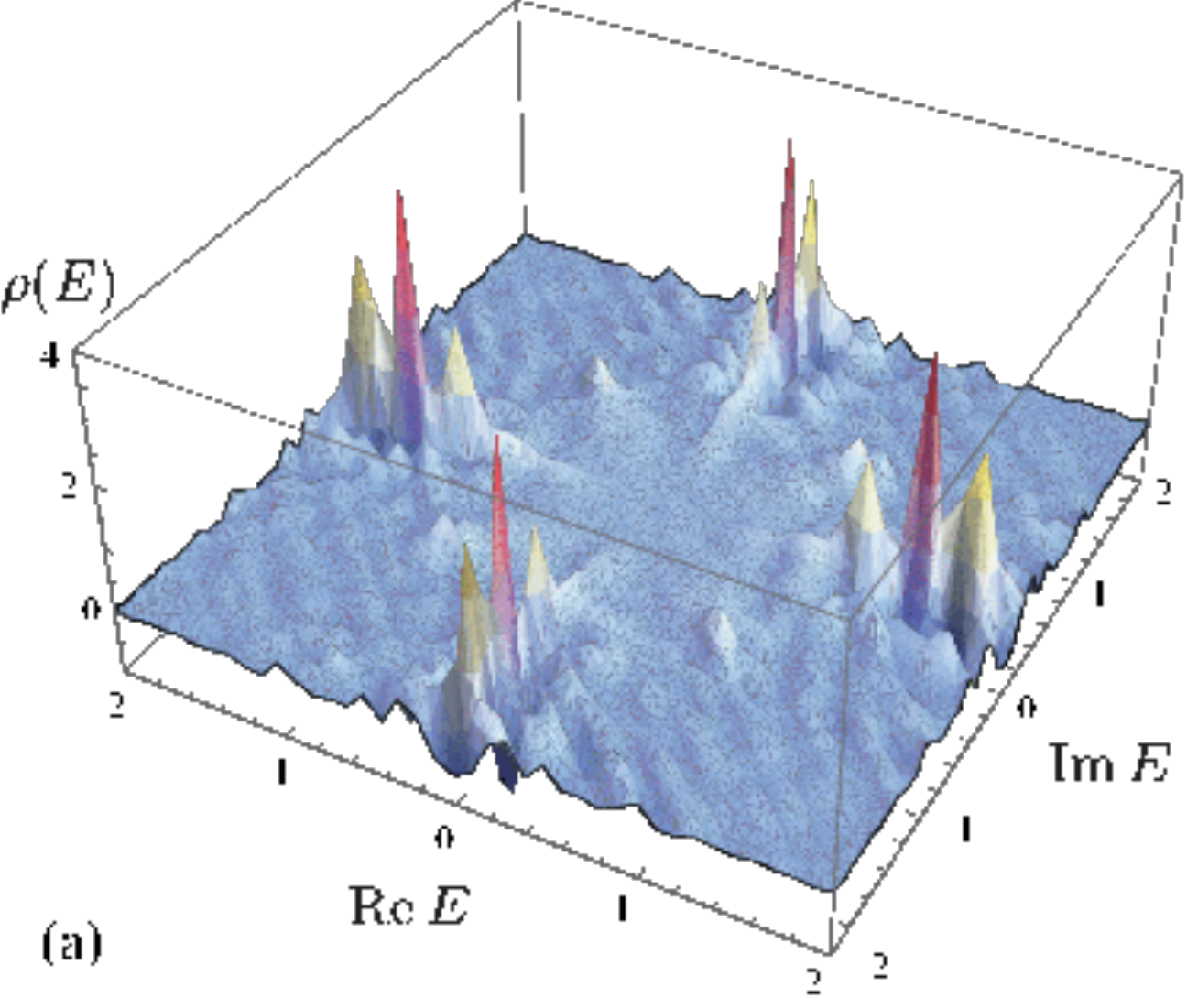}
\vspace{\baselineskip}
\\
\includegraphics[width=0.38\textwidth]{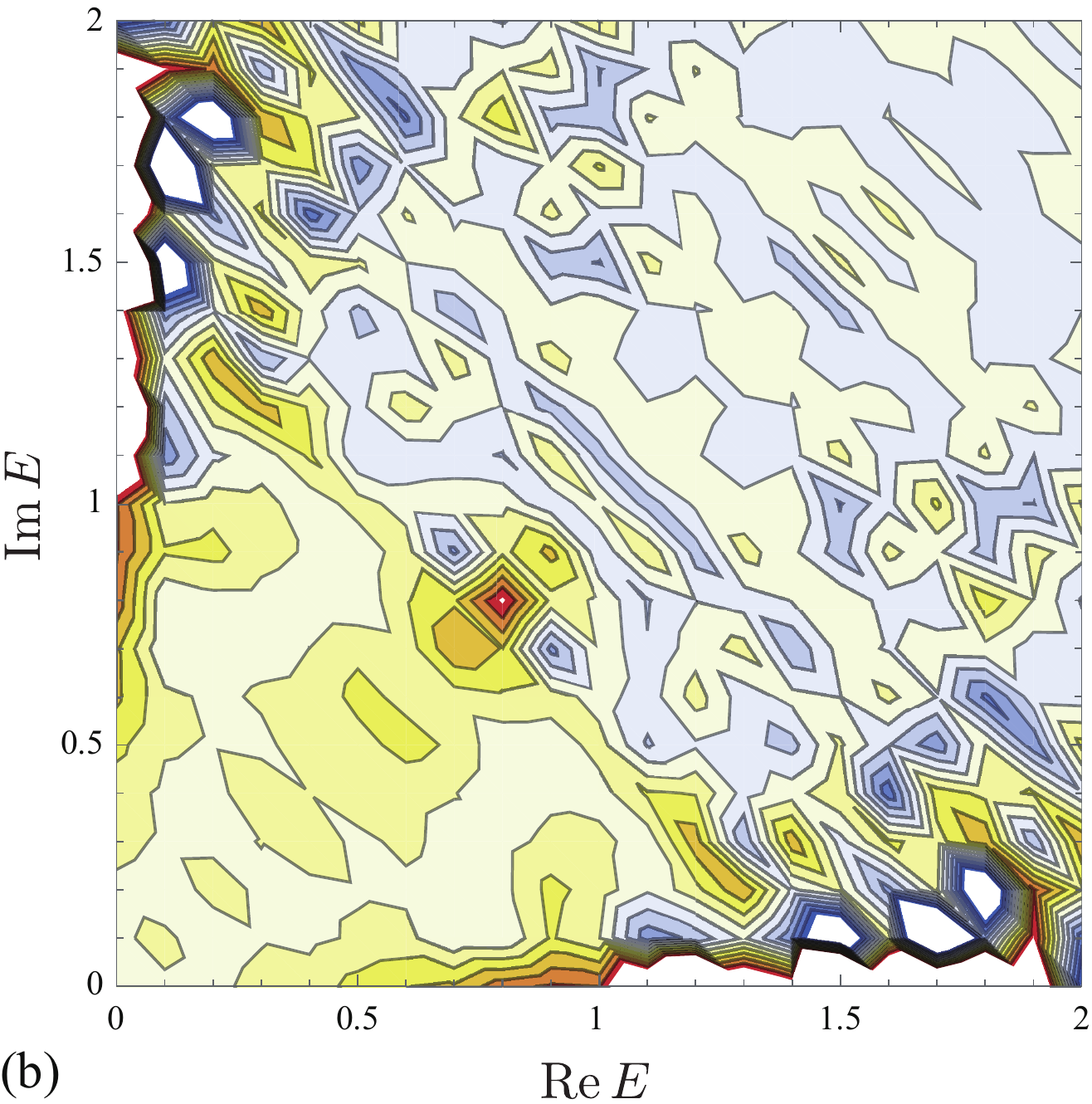}
\caption{(a) A three-dimensional plot of the density of states of 660 random samples of the FZ random-hopping model~\eqref{eq610} of length $500$, from the results of the expansion~\eqref{eq400-1} up to the $500$th order.
We used the data of the first quadrant to plot the other quadrants.
(b) A contour plot of the first quadrant of the same data.
The level contours indicate the data from $-0.5$ to $0.6$ in increments of $0.1$.
The peaks around the real and imaginary axes are cut off.
We have removed the normalization of the spectrum and plot the results in the original energy scale.}
\label{fig8}
\end{figure}

We shall demonstrate the expansion~\eqref{eq400} of $\rho(z,z^\ast)$ first for the FZ random-hopping model~\eqref{eq610} and second for correlated random-sign matrices.
Note the fractal-like structure of the spectrum in Fig.~\ref{fig5}~(a) of the FZ random-hopping model.
Clearly, truncating the series~\eqref{eq400-1} can only be expected to reproduce a coarse-grained approximation to its finely featured spectrum;
if it is truly fractal, and hence singular, we will never be able to express it in terms of a finite-order polynomial.
This is indeed what we observe; see Fig.~\ref{fig8}.
Although the peaks on the real and imaginary axes are consistent with the numerical results in Fig.~3 of Ref.~\cite{AHN15}, we can have only a rough idea of the spectrum in between.

Next, as promised,  we apply the expansion~\eqref{eq400} for a non-Hermitian model with dense and smooth spectrum.
We draw a full  $L\times L$ random matrix $H$ from an ensemble as follows;
for each pair of off-diagonal elements $H_{ij}$ and $H_{ji}$, we set both of them equal to $\pm1$ with probability $\tau$ (that is,  to $+1$ with probability $\tau/2$ and to $-1$ with probability $\tau/2$), but set them independently randomly to $\pm1$ with probability $1-\tau$, while setting all diagonal elements to zero.
This means a partially symmetric real random matrix with the correlation $\langle H_{ij}H_{ji}\rangle=\tau$.

\begin{figure*}
\includegraphics[width=0.4\textwidth]{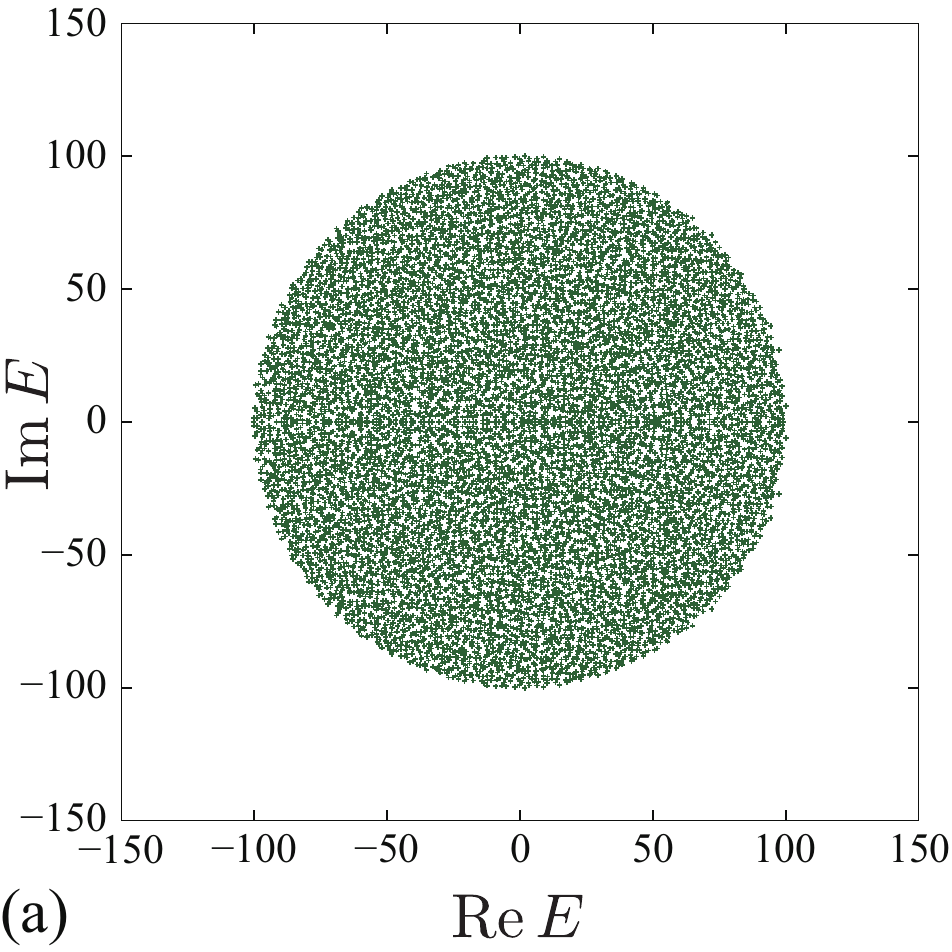}
\hspace{0.06\textwidth}
\includegraphics[width=0.4\textwidth]{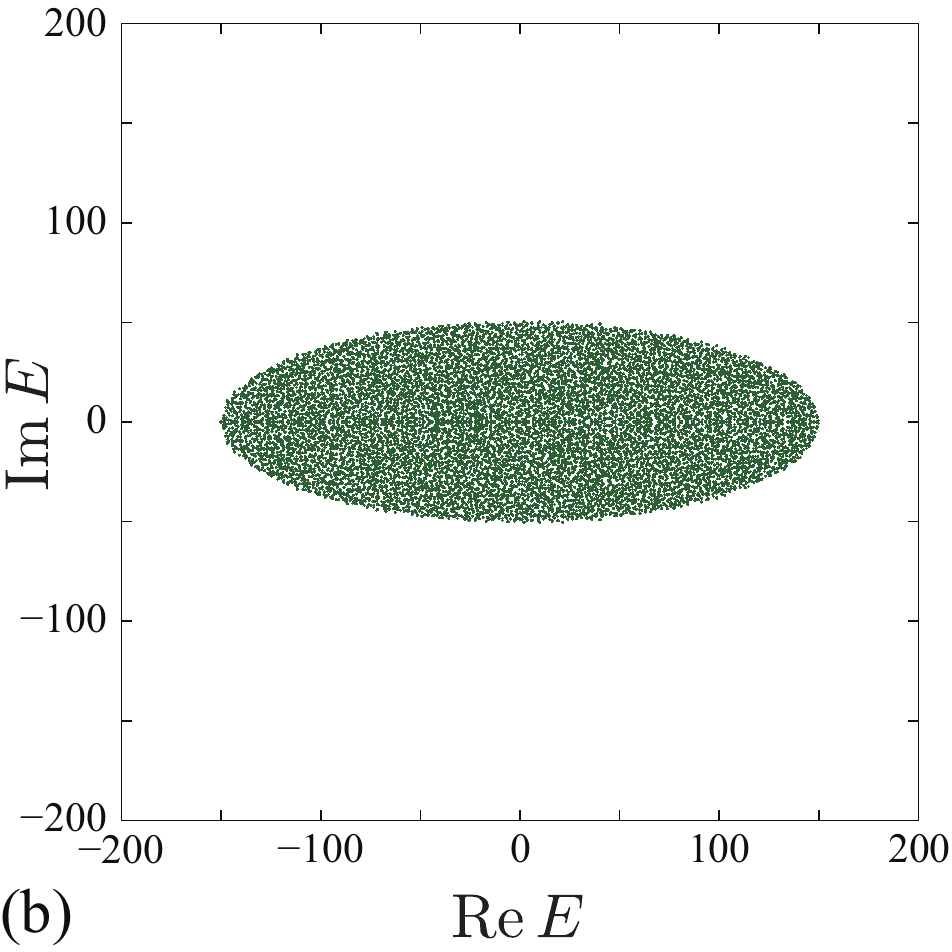}
\vspace{\baselineskip}
\\
\includegraphics[width=0.42\textwidth]{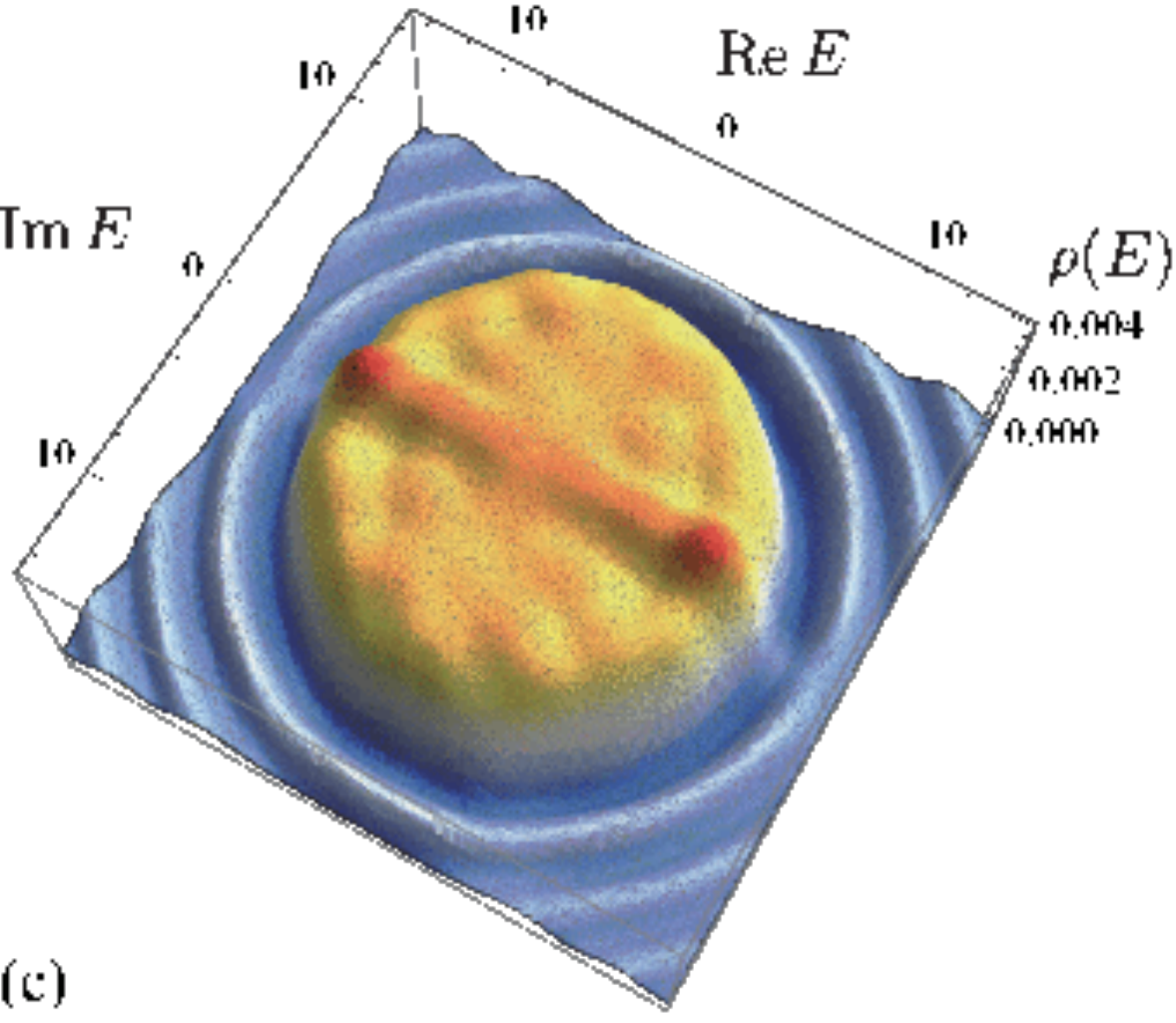}
\hspace{0.05\textwidth}
\includegraphics[width=0.42\textwidth]{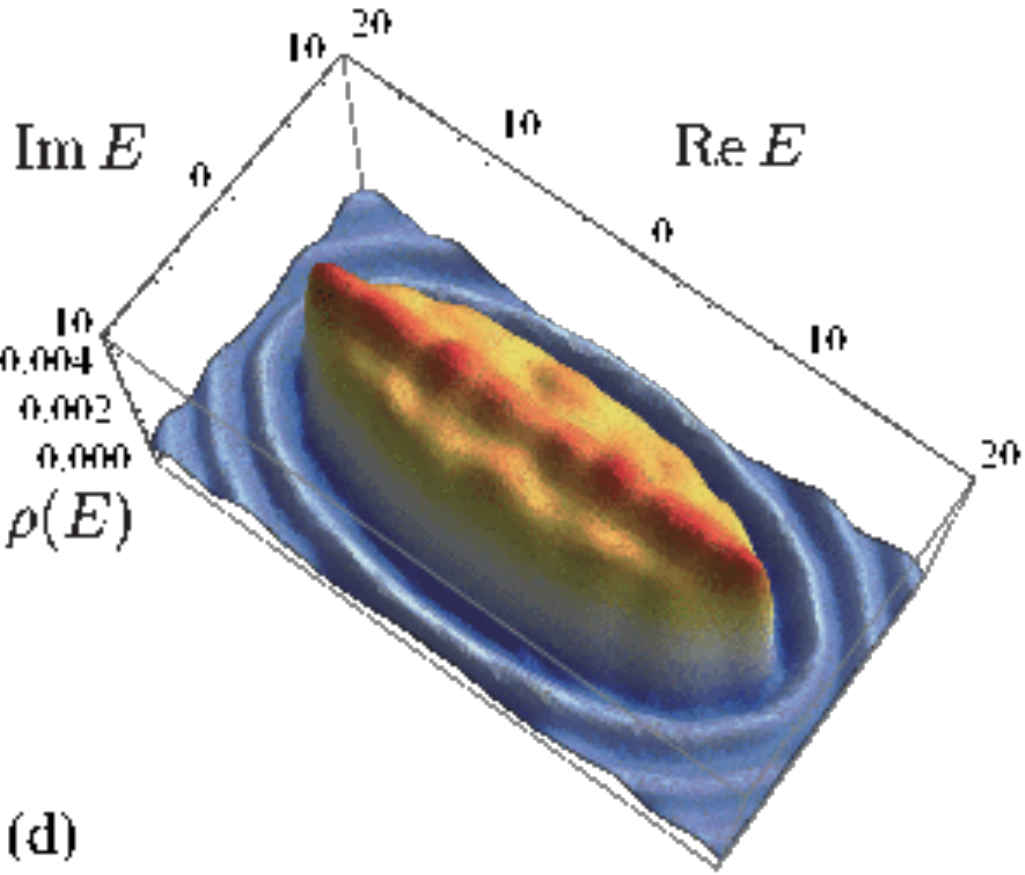}
\caption{Plots of the eigenvalue distributions of (a) the fully asymmetric random matrix ($\tau=0$) and (b) a partially asymmetric random matrix ($\tau=0.5$), both of size $10000\times10000$, obtained by means of diagonalization, together with three-dimensional plots of the density of states of 96 random samples of (c) the fully asymmetric random matrix ($\tau=0$) and (d) a partially asymmetric random matrix ($\tau=0.5$), both of size $100\times100$, from the results of the expansion~\eqref{eq400-1} up to the $500$th order.
Here we have removed the normalization of the spectrum and plot the results in the original energy scale.}
\label{fig9}
\end{figure*}

According to Ref.~\cite{Sommers88} for Gaussian randomness (consistent with $\langle H_{ij}H_{ji}\rangle=\tau$), the density of states is uniform inside an ellipse:
\begin{align}\label{eq690}
\rho(E, E^*)=\begin{cases}
(\pi a b)^{-1}& \mbox{if $({\rm Re}E/a)^2+({\rm Im}E/b)^2\leq 1$,} \\
0 & \mbox{otherwise,}
\end{cases}
\end{align}
where $a=\sqrt{L}(1+\tau)$ and $b=\sqrt{L}(1-\tau)$.
This reduces to the celebrated Wigner semi-circle law~\cite{Mehta04} on the real axis in the completely symmetric case, namely $\tau=1$, and to Girko's circle law~\cite{Girko85} in the completely asymmetric case, namely, $\tau=0$.
Figures~\ref{fig9}~(a) and~(b) show the results of the diagonalization of $10000\times10000$ random-sign matrices with $\tau=0$ and $\tau=0.5$, respectively, which are  indeed consistent with the law~\eqref{eq690} for Gaussian random matrices, thus demonstrating universal behavior.
Our expansion~\eqref{eq400-1}, on the other hand, produces Fig.~\ref{fig9}~(c) and~(d) for $100\times100$ random matrices.
We can see a ridge on the real axis, which is  a finite-size effect.

For relatively small full matrices, such as the matrices in the present demonstration (Fig.~\ref{fig9}), the expansion~\eqref{eq400} for the density of states is a much more time-consuming algorithm than direct numerical diagonalization.
For large general non-Hermitian matrices, on the other hand, the expansion ~\eqref{eq400} would be almost the only available algorithm at reasonable CPU cost. Moreover, since it is of the form of repeated multiplication of a vector by a matrix, it can be quite efficient for sparse matrices, which appear often in many-body systems with interaction.
For large dense matrices, we might need to replace the trace operation in Eq.~\eqref{eq400} with Monte Carlo summation over a set of basis vectors less than $L$.

In conclusion of this section, let us briefly mention again the works in Refs.~\cite{Rogers09,Metz11,Neri12}, which calculated, both analytically and numerically, spectra of large sparse non-Hermitian matrices \textit{of certain types} by alternative methods. 
All these works made use of the method of Hermitization described in Sec.~\ref{sec3.1}; the first one~\cite{Rogers09}, in particular, analyzed as a test case the same matrix model as in the latter example above.

\section{Summary}
\label{sec5}

We have derived the Chebyshev-polynomial expansion of the energy dependence of the inverse localization length both for Hermitian and non-Hermitian chains.
For Hermitian chains, the expansion produces the energy dependence as a function in one run.
This is in strong contrast to the standard transfer-matrix method, which produces the inverse localization at a fixed energy in one run.
Since our method is based on the repeated multiplication of some vector by a Hamiltonian matrix, which is sparse in many cases, we can, in principle,  execute the algorithm by storing only vectors; 
we do not need to store the entire matrix in the computer memory.
This feature may come in handy when we try to generalize the method to models with interactions in the future.
Although the Thouless formula does not apply, at least directly, to interacting systems, the ``localization length" (that is, the length scale governing the decay) of the two-point Green's function is still an important concept in studying \textit{e.g.}\ the Anderson-Mott transition~\cite{Belitz94}.

For non-Hermitian chains, the expansion produces only the inverse localization length at a fixed (complex) energy in one run.
Again, we do not need to store the entire matrix in computer memory.
We have also found the Chebyshev-polynomial expansion of the density of states for non-Hermitian matrices.
The present algorithm may be almost the only available algorithm of finding the density of states without storing the entire matrix in computer memory.

Yet another important application of the method in studying the Anderson localization is to compute the density of resonances of open disordered systems, particularly in three dimensions. 
We can find the resonances as eigenvalues of an effective non-Hermitian Hamiltonian, which we obtain from the full Hamiltonian, describing the system of interest and the environment it is coupled to, after properly eliminating the latter and the outgoing waves in it as described \textit{e.g.}\ in Refs.~\cite{Kunz06,Kunz08,Feinberg09,Feinberg11,Sasada08,Sasada11}.
The density of resonances of an open disordered three-dimensional system, unlike the density of states of a closed system, can distinguish the localized and extended phases on the both sides of the Anderson transition in higher dimensions. See \textit{e.g.}\ Ref.~\cite{Kottos05}.


\appendix

\section{Integration of Equation~\eqref{eq170}}
\label{appA}

In order to find the integral $f_n(E)$ in Eq.~\eqref{eq170}, we first differentiate it with respect to $E$:
\begin{align}
f_n'(E)=\int_{-1}^1\!\!\!\!\!\!\!\!\!-       \hspace{0.4cm}\frac{T_n(x)}{(E-x)\sqrt{1-x^2}}dx\,,\quad\quad |E|<1.
\end{align}
We can find the formulae for this integral in 7.344.1 of Ref.~\cite{GR}; the answer is
\begin{align}\label{eqA190}
f_n'(E)=-\pi U_{n-1}(E)
\end{align}
for $n>0$, where $U_n(x)$ is the Chebyshev polynomial of the second kind.
For $n=0$, we have
\begin{align}\label{eqA200}
f_0'(E)
&=\lim_{\varepsilon\to0}\left(\int_{-1}^{E-\varepsilon} +\int_{E+\varepsilon}^1\right) \frac{1}{(E-x)\sqrt{1-x^2}}dx
=0.
\end{align}
Equation~\eqref{eqA190} is followed by
\begin{align}\label{eqA230}
f_n(E)=-\pi \int^E U_{n-1}(x)dx=-\frac{\pi}{n}T_n(E)+\mathrm{const}.
\end{align}
for $n>0$ since  
\begin{align}
\frac{d}{dx}T_n(x)=nU_{n-1}(x).
\end{align}
Equation~\eqref{eqA200} gives
\begin{align}\label{eqA250}
f_0(E)=\mathrm{const}.
\end{align}

Finally, we can fix the constants in Eqs.~\eqref{eqA230} and~\eqref{eqA250} as follows.
First, we have
\begin{align}
f_0(1)&=\int_{-1}^1\frac{\ln(1-x)}{\sqrt{1-x^2}}dx=-\pi\ln2,
\\
f_1(1)&=\int_{-1}^1\frac{x\ln(1-x)}{\sqrt{1-x^2}}dx=-\pi;
\end{align}
see 4.292.3 and 4.292.4 of Ref.~\cite{GR}, respectively.
Further integrations can be obtained from Eq.~(2.10) of Ref.~\cite{Boyd89}:
\begin{align}\label{eqA90}
\ln(1-x)=-\ln 2 - 2\sum_{n=1}^\infty \frac{T_n(x)}{n}.
\end{align}
We therefore have
\begin{align}\label{fn1}
f_n(1)&=\int_{-1}^1\frac{T_n(x)\ln(1-x)}{\sqrt{1-x^2}}dx=-\frac{\pi}{n}
\end{align}
for $n>0$.
We thereby conclude that
\begin{align}\label{fnE}
f_0(E)&=-\pi\ln2,
\\
f_n(E)&=-\frac{\pi}{n}T_n(E)
\end{align}
for $n>0$, where we used $T_n(1)=1$.

\section{Structure of the recursion relation~\eqref{eq340}}
\label{appB}

We here show the matrix structure of the recursion relation~\eqref{eq340}.
Throughout this appendix, we use the shorthand notation $A=H-z$ and $A^\dag=H^\dag-z^\ast$.

It is easy to prove inductively that the even-order matrix polynomials $T_{2m}(\H)$ have non-vanishing elements only in the  $L\times L$ upper-left and lower-right diagonal blocks, which we denote by $T_{2m}^{(1,1)}$ and $T_{2m}^{(2,2)}$, respectively, while the odd-order ones $T_{2m+1}(\H)$ have non-vanishing elements only in the upper-right and lower-left blocks, which we denote by $T_{2m+1}^{(1,2)}$ and  $T_{2m+1}^{(2,1)}$, respectively.
The recursion relation~\eqref{eq340} indeed reads
\begin{align}\label{eqB790}
\begin{pmatrix}
T_{2m}^{(1,1)} & 0 \\
0 & T_{2m}^{(2,2)}
\end{pmatrix}
&=
2
\begin{pmatrix}
0 & A \\
A^\dag & 0
\end{pmatrix}
\begin{pmatrix}
0 & T_{2m-1}^{(1,2)} \\
T_{2m-1}^{(2,1)} & 0
\end{pmatrix}
\nonumber\\
&-
\begin{pmatrix}
T_{2m-2}^{(1,1)} & 0 \\
0 & T_{2m-2}^{(2,2)}
\end{pmatrix},
\\\label{eqB800}
\begin{pmatrix}
0 & T_{2m+1}^{(1,2)} \\
T_{2m+1}^{(2,1)} & 0
\end{pmatrix}
&=
2
\begin{pmatrix}
0 & A \\
A^\dag & 0
\end{pmatrix}
\begin{pmatrix}
T_{2m}^{(1,1)} & 0 \\
0 & T_{2m}^{(2,2)}
\end{pmatrix}
\nonumber\\
&-
\begin{pmatrix}
0 & T_{2m-1}^{(1,2)} \\
T_{2m-1}^{(2,1)} & 0
\end{pmatrix},
\end{align}
which obviously prove the point.

The explicit forms of the  first few matrix polynomials are
\begin{align}
T_0(\H)&=\begin{pmatrix}
I & 0 \\
0 & I
\end{pmatrix},
\\
T_1(\H)&=\begin{pmatrix}
0 & A \\
A^\dag & 0
\end{pmatrix},
\\
T_2(\H)&=\begin{pmatrix}
2AA^\dag-I & 0 \\
0 & 2A^\dag A-I
\end{pmatrix},
\\
T_3(\H)&=\begin{pmatrix}
0 & A(4A^\dag A-3) \\
A^\dag(4AA^\dag-3) & 0
\end{pmatrix},
\\
T_4(\H)&=\begin{pmatrix}
8(AA^\dag)^2-8AA^\dag+I & 0 \\
0 & 8(A^\dag A)^2-8A^\dag A+I
\end{pmatrix}\,.
\end{align}
Let us presume that 
\begin{align}\label{eqB860}
T_{2m}^{(1,1)}&=T_{2m}\left(\sqrt{AA^\dag}\right),\\
T_{2m}^{(2,2)}&=T_{2m}\left(\sqrt{A^\dag A}\right),\\
T_{2m+1}^{(1,2)}&=A\tilde{T}_{2m+1}\left(\sqrt{A^\dag A}\right),\\\label{eqB890}
T_{2m+1}^{(2,1)}&=A^\dag\tilde{T}_{2m+1}\left(\sqrt{AA^\dag}\right),
\end{align}
where we used a temporary notation $\tilde{T}_{2m+1}(x)=x^{-1}T_{2m+1}(x)$.
It is indeed easy to prove this inductively by inserting Eqs.~\eqref{eqB860}--\eqref{eqB890} into the recursion relations~\eqref{eqB790} and~\eqref{eqB800}.

We thereby conclude that 
\begin{align}\label{eqB865}
\frac{1}{2L}\tr T_{2m}(\H)&=\frac{1}{L}\tr T_{2m}^{(1,1)}=\frac{1}{L}\tr T_{2m}^{(2,2)},
\\
\frac{1}{2L}\tr T_{2m+1}(\H)&=0\,.
\end{align}

\section{Integration of Equation~\eqref{eq370}}
\label{appC}

In view of the representation $T_n(x)=\cos(n\arccos x)$ [Eq.~\eqref{eq20}] for Chebyshev polynomials, we rewrite the left-hand side of Eq.~\eqref{eq370} as
\begin{align}\label{a20}
I_{2m}=\int_{0}^{\pi/2}\cos(2m\theta)\ln(\cos\theta) d\theta.
\end{align}

For $m>0$, we integrate by parts and obtain
\begin{align}
I_{2m}&=\frac{1}{2m}\left[\sin(2m\theta)\ln\cos\theta\right]_{\theta=0}^{\pi/2}
\nonumber\\
&+\frac{1}{2m}\int_{0}^{\pi/2}\sin(2m\theta)\frac{\sin\theta}{\cos\theta}d\theta.
\end{align}
The boundary term clearly vanishes. 
We thus have
\begin{align}
I_{2m}=-\frac{1}{2m}\int_{0}^{\pi/2}\frac{\cos[(2m+1)\theta]-\cos[(2m-1)\theta]}{2\cos\theta}d\theta.
\end{align}
We know this integration from the formula
\begin{align}
\int_0^\pi
\frac{\sin[(2m+1)\theta]}{\sin\theta}d\theta
=\pi
\end{align}
independently of $m$.
We therefore have
\begin{align}
&\int_{0}^{\pi/2}\frac{\cos[(2m\pm1)\theta]}{\cos\theta}d\theta
\nonumber\\
&=\pm(-1)^m\int_{0}^{\pi/2}\frac{\sin[(2m+1)\theta]}{\sin\theta}d\theta
=\pm(-1)^m\frac{\pi}{2},
\end{align}
and arrive at the final formula
\begin{align}
I_{2m}=-\frac{\pi}{2}\frac{(-1)^m}{2m}.
\end{align}

\section{Equations~\eqref{eq210} and~\eqref{eq410} for Hermitian matrices}
\label{appD}


We here show that the Chebyshev expansion~\eqref{eq410} for non-Hermitian matrices reduces to the expansion~\eqref{eq210} when $H$ is a Hermitian matrix.
For the purpose, we first introduce summation formulas for Chebyshev polynomials.
We start from the Fourier series
\begin{align}\label{eqD1000}
\sum_{m=1} (-1)^m \frac{\cos 2m\theta}{m}
=-\ln \left| 2\cos \theta \right|.
\end{align}
Setting $\theta=\arccos(x-y)$ in Eq.~\eqref{eqD1000} to use the definition $T_n(x)=\cos(n \arccos x)$, we have
\begin{align}\label{eqD1005}
\sum_{m=1}^\infty (-1)^m \frac{T_{2m}(x-y)}{m}=-\ln 2|x-y|.
\end{align}
We then next use the formulas
\begin{align}\label{eqD1010}
\sum_{n=1}^\infty \frac{\cos n\theta \cos n\phi}{n}&=-\frac{1}{2}\ln 2|\cos\theta-\cos\phi|,
\\\label{eqD1020}
\sum_{n=1}^\infty (-1)^n \frac{\cos n\theta \cos n\phi}{n}&=-\frac{1}{2}\ln 2|\cos\theta+\cos\phi|,
\end{align}
which we can prove as follows:
\begin{widetext}
\begin{align}
\sum_{n=1}^\infty \frac{\cos n\theta \cos n\phi}{n}
&=\frac{1}{2}\sum_{n=1}^\infty \frac{\cos n(\theta+\phi)}{n}
+\frac{1}{2}\sum_{n=1}^\infty \frac{\cos n(\theta-\phi)}{n}
\nonumber\\
&=-\frac{1}{2}\ln\left|2\sin\frac{\theta-\phi}{2}\right|
-\frac{1}{2}\ln\left|2\sin\frac{\theta+\phi}{2}\right|
=-\frac{1}{2}\ln\left|4\sin\frac{\theta+\phi}{2}\sin\frac{\theta-\phi}{2}\right|
\nonumber\\
&=-\frac{1}{2}\ln | 2(\cos\theta-\cos\phi)|,
\\
\sum_{n=1}^\infty (-1)^n \frac{\cos n\theta \cos n\phi}{n}
&=\frac{1}{2}\sum_{n=1}^\infty (-1)^n \frac{\cos n(\theta+\phi)}{n}
+\frac{1}{2}\sum_{n=1}^\infty (-1)^n \frac{\cos n(\theta-\phi)}{n}
\nonumber\\
&=-\frac{1}{2}\ln\left|2\cos\frac{\theta-\phi}{2}\right|
-\frac{1}{2}\ln\left|2\cos\frac{\theta+\phi}{2}\right|
=-\frac{1}{2}\ln\left|4\cos\frac{\theta+\phi}{2}\cos\frac{\theta-\phi}{2}\right|
\nonumber\\
&=-\frac{1}{2}\ln | 2(\cos\theta+\cos\phi)|.
\end{align}
\end{widetext}
We set $\theta=\arccos x$ and $\phi=\arccos y$ in Eqs.~\eqref{eqD1010} and~\eqref{eqD1020} this time, having
\begin{align}\label{eqD1050}
\sum_{n=1}^\infty \frac{T_n(x)T_n(y)}{n}&=-\frac{1}{2}\ln 2|x-y|,
\\\label{eq1060}
\sum_{n=1}^\infty (-1)^n \frac{T_n(x)T_n(y)}{n}&=-\frac{1}{2}\ln 2|x+y|.
\end{align}
Comparing Eqs.~\eqref{eqD1005} and~\eqref{eqD1050}, we arrive at the formula
\begin{align}\label{eqD1060}
\sum_{m=1}^\infty (-1)^m \frac{T_{2m}(x-y)}{m}=2\sum_{n=1}^\infty \frac{T_n(x)T_n(y)}{n}.
\end{align}

We use the formula~\eqref{eqD1060} to prove that Eq.~\eqref{eq410} reduces to Eq.~\eqref{eq210} for a Hermitian matrix $H$.
We here make use of the last form of Eq.~\eqref{eqB865} for the expansion coefficient in Eq.~\eqref{eq410}.
For Hermitian matrices, the eigenvalues $\{E_\nu\}$ of $H$ are all real.
We put $z$ to the real variable $E$, because we are interested in $\kappa(E)$ on the real axis in Eq.~\eqref{eq210}.
We can therefore cast Eq.~\eqref{eq410} into the form
\begin{align}\label{eqD960}
\kappa(E)=-\frac{1}{L}\sum_{\nu=1}^L
\sum_{m=1}^\infty
\frac{(-1)^m}{m}
T_{2m}(E_\nu-E)
-\ln (2|\tau|).
\end{align}
We are now in a position to employ the formula~\eqref{eqD1060} to transform Eq.~\eqref{eqD960} into
\begin{align}
\kappa(E)&=-\frac{2}{L}\sum_{\nu=1}^L
\sum_{n=1}^\infty
\frac{T_n(E_\nu)T_n(E)}{n}
-\ln(2 |\tau|),
\end{align}
which is indeed equal to Eq.~\eqref{eq210}.

\begin{acknowledgments}
NH greatly appreciates the hospitality of Department of Physics, Technion, and particularly Prof.\ Dov Levine for the support of the stay.
NH's research is partially supported by Kakenhi Grants No. 15K05200, No. 15K05207, and No. 26400409 from Japan Society for the Promotion of Science.
\end{acknowledgments}

\bibliography{hatano}

\end{document}